\documentclass[11pt,letterpaper]{article}

\usepackage{graphicx}
\usepackage{natbib}
\usepackage{amsmath}
\usepackage{float}
\usepackage{subfig}
\usepackage{booktabs}
\usepackage{caption}
\usepackage[flushleft]{threeparttable}
\usepackage{longtable}
\usepackage{geometry}
\usepackage{setspace}

\makeatletter
\renewcommand{\@makefntext}[1]{%
  \noindent
  \parindent 0pt
  \makebox[0.8em][l]{\@makefnmark}#1}
\makeatother

\newcommand{\titlerunning}[1]{}

\begin{document}

\title{Gas Giant and Brown Dwarf Companions: Mass Ratio and Orbital Distributions From A stars to M dwarfs}

\author{
  Michael R. Meyer$^{*,1}$ \and
  Yiting Li$^{*,1}$ \and
  Per Calissendorf$^{1}$ \and
  Adam Amara$^{2}$
}

\footnotetext[1]{Department of Astronomy, The University of Michigan; \\$^{2}$School of Physics and Math, The University of Surrey;\\ $^{*}$These authors contributed equally to this work.}

\date{Submitted to the A\&A.}

\maketitle
\nopagebreak

\begin{abstract}
Understanding demographic properties of planet populations and multiple star systems constrains theories of planet and star formation. Surveys for very low-mass companions to M-A type stars detect brown dwarfs from multiple star formation and planets from circumstellar disks. We fit a composite model describing both very low-mass brown dwarf companions from ``multiple-like processes" and gas giants from ``planet-like processes" as functions of orbital separation and host star mass. We assemble a database of companion frequency estimates for masses from $<$ 1 to $>$ 75 Jupiter masses, separations from $<$ 0.3 to $>$ 300 AU, and host masses from $<$ 0.3 to $>$ 2 $M_{\odot}$. Using multinest, we fit these data to various models, performing model selection and deriving probability density functions. We assume companion mass ratio distributions are independent of orbital separation and fit a common log-normal orbital distribution to gas giant populations around M dwarfs, FGK, and A stars. A six-parameter model based on companion mass ratio distributions for planets and brown dwarfs is preferred. The planet CMRD slope is consistent with previous studies ($dN/dq \sim q^{-1.3} \pm 0.03$). Gas giant planets around stars from $<$ 0.3 to $>$ 2.0 $M_{\odot}$ follow a log-normal distribution peaking at ln(a) = 1.30 $\pm$ 0.03 (3.8 AU) with dispersion 0.22 $\pm$ 0.04. M dwarf distributions peak at smaller orbital radii than A stars, consistent with iceline considerations. Brown dwarf companion distributions extend stellar binary patterns, with the brown dwarf desert explained by flat-in-q mass functions and limited mass ratios below 0.1.
\end{abstract}


\section{Introduction}
   Isolated stars that are not part of a multiple system \citep{2013ARA&A..51..269D}, and lack planetary systems \citep{2013ApJ...770...69P}, are rare in the Milky Way (and presumably other galaxies).  Apparently, nature disdains physical isolation.  Multiple star systems can form through fragmentation of turbulent molecular cores, the breakup of flattened circumstellar structures that are only partially supported by rotation, and gravitational instability in a fully Keplerian disk (\cite{2010ApJ...725.1485O}; \cite{2016ARA&A..54..271K}).  Each channel of formation may contribute differently as a function of the mass ratio of the companion to central star mass, orbital separation, and total system mass.  Gas giant planets can also form through gravitational instability \citep{2018ApJ...854..112M}.  However, it is likely that core accretion, with phases of core formation and rapid gas accumulation from the primordial disk, dominates the formation of companions below three Jupiter masses \citep[cf.][]{2014prpl.conf..691B}.  Fragmentation in the interstellar medium, and perhaps circumstellar disks, is likely limited to $>$ the opacity limit estimated to be $<$ 10 Jupiter masses \citep[e.g.][]{1978ppim.book.....S}. Gravitational instability (and perhaps core accretion) can form objects in principle up to 10s of Jupiter masses leaving a significant range of overlap in mass between objects formed from “multiple-like” versus “planet-like” processes.  Yet during formation and early evolution, neither class of formation pathway is aware of the deuterium burning limit at 13 Jupiter masses, which the IAU has adopted as the boundary between a “planet” and a “brown dwarf”.  Furthermore, both kinds of objects may be detected in surveys for very low mass companions from $<$ 1 AU (e.g. radial velocity and transit; ), 1-10 AU (e.g. long-term RV, microlensing, and astrometry; \cite{2016ApJ...819...28W}; \citet{2016ApJ...833..145S, 2011A&A...525A..95S}), and beyond 10 AU (e.g. direct imaging and astrometry from the Gaia mission; \citet{2016A&A...586A.147R}; \cite{2018A&A...614A..30R}).  Properly accounting for the contribution of brown dwarf companions and gas giants to mass ranges of overlap requires a model that describes both.  How can we tell the difference between planets and brown dwarfs?  There is some evidence from analysis of radial velocity samples of discontinuities in the mass function as a function of host star metallicity (\citet{2015A&A...576A..94S, 2018ApJ...853...37S}).   \citet{2020AJ....159...63B} also find a dichotomy between the eccentricity distribution of brown dwarf-like populations (more like stellar multiples with a uniform eccentricity distribution) and planets (lower average eccentricities, suggesting formation in a Keplerian disk).  \cite{2023AJ....165..164B} likewise find hints of a difference in the distribution of orbital obliquities between lower mass close-in companions (presumably more ``planet-like") and more massive wide-orbit companions (presumably more ``brown-dwarf like").  Another approach is to compare the composition of planetary mass companions in key volatile species to those of the host stars.  Companions strongly enriched in heavy elements (like Jupiter) would indicate formation in a circumstellar disk \citep[e.g.][]{2019ARA&A..57..617M} see also \citet{2023AJ....166..192A}. \par

    We seek to explore the differences between brown dwarfs and planetary mass companions, in order to make testable predictions regarding very low mass companion frequencies, and provide relative population constraints to inform star and planet formation theory.  To do this, we have assembled a database of point estimates of companion frequency from the literature, as a function of companion mass ratio, orbital separation, and host star mass.  We then fit the collected occurrence rates versus orbital separation and mass ratio distributions to various functional forms to find the best--fitting model and probability density functions (PDFs) of the fit variables.  Some assumptions are required.  For example, 
    we assume that companion mass functions do not depend on orbital separation and provide justification for this below. 
    The frequency of stellar companions depends on orbital separation in ways that differ from gas giant planets for M, FGK, and A stars. We explore whether brown dwarf companions follow similar orbital distributions.  
    In section 2 we present our methodology.  In section 3, we report the results of our model fitting.  In section 4 we discuss the implications of our model for star and planet formation theory.  In section 5, we summarize our results and suggest future work. \par

\section{Methodology}

    We fit our models to point estimates of companion frequency, as a function of the range of masses, and mass ratios ($q = M_{companion}/M_{star}$) and orbital separations surveyed.  We describe the parameters of our models, the database assembled, and then the fitting procedure used. 
  
  \subsection{The Models}
  
Our models are comprised of two components: one describes the gas giant planet population from $<$ 1 Jupiter mass to $>$ 10 Jupiter mass from $<$ 0.3 AU to $>$ 300 AU and another describes the brown dwarf companion frequency over the same parameter space.  We fit all available data for primaries from $<$ 0.3 to $>$ 2.0 M$_{\odot}$ simultaneously, although we also explicitly search for a stellar mass dependence in the orbital separation of planets and companion frequency.  If we assume that both populations contribute to an observed total number of companions, and that the mass distributions of each are independent of the respective orbital separation distributions, we can write:
  
    $$N_{TOTAL} = \int{\phi_{p}(x) \times \psi_{p}(q) dq dx} + \int{\phi_{bd}(x) \times \psi_{bd}(q) dq dx}$$  
    
where the first term is the planet contribution and the second term is the brown dwarf companion population.  

In our models, $\phi_{p}(x) = A_{p} e^{-(x - \mu_p)^2/2 \sigma_p^2}/x \sqrt{2\pi} \sigma_p$, where $x = ln(a)$, is the approximate physical (not projected) observed orbital separation distribution of planets, $A_p$ is the planet normalization coefficient, $\mu_p$ is the mean of the planet orbital separation distribution in ln(a), and $\sigma_p$ is the dispersion in units of ln(a).  We fit in the natural log, but also quote the results in log-10 below.  We fit a log-normal for the orbital separation distribution for all planetary mass companions, for all primaries\footnote{We tried to fit the orbital distributions for each host star mass group separately, but found that they were consistent with each other except as discussed below.}.  The three free parameters in this part of the model include the normalization coefficient, the mean, and the dispersion of the log-normal.

We also fit the global normalization of the brown dwarf companion frequency, for all host star masses, assuming the orbital separation distributions found for stellar mass companions as a function of host mass.  We adopt the log-normal fit to the orbital separation distribution of companions around M dwarfs from \citet{2019AJ....157..216W} adjusting from projected separation (20 AU) to physical separation (27 AU) with a correction factor of $\times$ 1.35 from \cite{1991A&A...248..485D}; hereafter DM91.  For FGK stars, we adopt the log-normal fit of \cite{2010ApJS..190....1R} with a peak in physical separation at 50 AU. \cite{2014MNRAS.437.1216D} present a log-normal fit for companions $>$ 30 AU surrounding A star hosts and we adjust the results from projected separation to physical separation using DM91 (physical peak at 522 AU corresponding to a projected peak at 387 AU). \footnote{We note that expected corrections from projected to physical separation depends on the assumed distribution of eccentricities.  We use the factor of 1.35 from DM91 to be consistent with \cite{2010ApJS..190....1R}.}.  
Parameters of these stellar orbital distribution functions over indicated ranges of q and physical separation for all stellar types are given in Table 1 (quoted in log-base-10 as in the original references).  We also tried to fit a log-normal orbital distribution for all brown dwarfs, distinct from the stellar companion distributions without success as discussed below.  

   \begin{table}[h]
    \begin{threeparttable}
        \caption{Companion Frequency (CF) \& Log--Normal Separation Distribution vs. Host Type}
        \label{table:1}
        \centering
        \begin{tabular}{ c c c c c c }
            \hline\hline
            Spectral Type & M$_{comp}$/M$_{star}$ & physical separation (AU) & CF & $\mu$ (base-10) & $\sigma$ (base-10) \\
            \hline 
          M & $0.6-1.0^a$ & $0-10,000^a$ & $0.236^a$ & $1.43^b$ & $1.21^b$ \\
          FGK & $0.1-1.0^c$ & $0-\infty^c$ & $0.61^c$ & $1.70^d$ & $1.68^d$ \\
          A & $0.1-1.0^e$ & $30-800^e$ & $0.219 \pm 0.026^e$ & $2.72^e$ & $0.79^e$ \\
            \hline 

        \end{tabular}
        \begin{tablenotes}
            \small
            \item References: \cite{2022A&A...657A..48S}$^a$; \cite{2019AJ....157..216W} corrected to \\ physical separation as discussed in the text$^b$; \cite{2016A&A...586A.147R}$^c$; and \\ \cite{2010ApJS..190....1R}$^d$; and \cite{2014MNRAS.437.1216D} corrected to physical separation$^e$.
        \end{tablenotes}
        \end{threeparttable}
    \end{table}
    
We reiterate our assumption that the companion mass ratio distributions are independent of orbital separation. RM13 tested whether there was any evidence that the CMRD of stellar companions depended on orbital separation between 0.1 and 100,000 AU and found none.  We note that \cite{2014MNRAS.437.1216D} report a dependence on orbital separation for stellar binaries, with a steeper power-law for companions beyond 125 AU \footnote{Most of our data described below is draw from studies that sample within 100 AU.}.  \cite{2010ApJS..190....1R} also comment that there is a modest excess of equal mass companions at small separations.  Finally, we note that \citet{2017ApJS..230...15M} find evidence that the companion mass function does vary with orbital separation for OB stars.  Further investigation of this issue is warranted and would provide strong constraints on the efficacy of various modes of multiple star formation.   As discussed below, we searched for evidence that the planetary mass ratio distribution depends on orbital separation and found none.  While the planet mass distribution within 0.1 AU may be different (e.g. the ratio of Hot Jupiters to smaller planets), there is no evidence that the companion mass ratio distribution $>$ 0.30 AU depends on orbital separation down to 30 Earth masses (\cite{2021ApJS..255....8R}).  However absence of evidence is not evidence of absence. 

The planetary mass ratio distribution $\psi_{p}$ is equal to $dN/dq \sim q^{-\alpha}$ and $\psi_{bd} \sim q^{-\beta}$ is the analogous function for brown dwarf companions, which may have a different power-law slope.  We fit for these two power-law indices in a variety of cases.  Previous work suggests that the planet mass function is consistent with a power-law of slope $\alpha \sim 1.3$ (\citet{2008PASP..120..531C}) while the stellar companion mass ratio distribution is nearly flat (e.g. $\beta \sim 0$; \citep{2013A&A...553A.124R}). 
   We adopt these power-law forms, but in the mass ratio (q) distribution, and fit for the exponents.  We consider limits up to q = 0.1 for planets (the maximum mass of a stable circumstellar disk around a young star; e.g. \cite{1989ApJ...347..959A}) and down to $<$ 0.1 Jupiter mass, assuming it describes planet populations around stars from $<$ 0.3 to $>$ 2 M$_{\odot}$.  For brown dwarf companions formed by other means, we consider q ranges from the opacity limit for fragmentation 
   ($\leq$ 10 Jupiter masses) up to that corresponding to 85 Jupiter masses (the upper limit of occurrence rates in our database) as a function of host star mass. 

To summarize, we have explored models between six and nine parameters.  Our fiducial model includes the log-normal planet normalization (A$_p$), the log-normal mean ($\mu_p$) and dispersion ($\sigma_p$), the planet mass ratio distribution power--law index ($\alpha$), the brown dwarf companion normalization (A$_{bd}$), the brown dwarf companion mass ratio power--law index ($\beta$).  We fitted various characterizations of the brown dwarf orbital distribution such as an independent log-normal distributions (with two additional parameters).  We also compare fits using a log--normal fit to the planet populations to an alternate form where the separation distribution is a power--law in semi-major axis, with a sharp outer cut--off radius (\cite{2016ApJ...819..125C}).  This model replaces $\mu_p$ and $\sigma_p$ with an orbital separation power--law exponent and cut--off, having the same numbers of parameters. 

   In most of our fits we fixed the brown dwarf companion mass cutoff to 3 Jupiter masses, suggested by recent studies of young embedded star clusters probed 
   deeply with JWST (\cite{2025ApJ...981L..34D}).  We tried fitting the cutoff as a parameter of the model but the improvement in the fit was marginal and the parameter was unconstrained.  We also explore a planet mass function where $\psi_{p} \sim (M_*/M_{\odot})^{\gamma} m_p^{-\alpha}$ fixing the brown dwarf mass ratio distribution in q.  These models fit for a power--law planet mass function but with a stellar mass dependence introduced as a factor in the normalization.  This is another way to account for the observed increase in gas giant planet frequency with stellar mass (\cite{2016PASP..128j2001B}).  In the mass ratio models, the increase in frequency with stellar mass is interpreted as the result of detectable gas giants being of lower q in the power-law distribution, but constant normalization.  In the other model, the increase in frequency of gas giants with host star mass is due to an explicit dependence of the normalization factor on stellar mass, which could be interpreted as increased gas giant planet formation efficiency. Note that if $\gamma = \alpha$, the latter models would reduce identically to the mass ratio model, where the stellar mass dependence in the planet mass function appears as the mass ratio, q.  We generally compared results of our seven parameter 
   fits to our fiducial six parameter fits, but we also fit some eight and nine parameter models with additional variables as described below. 
  
  \subsection{The Database}
  
    We have assembled a database of point estimates for the frequency of companions over a fixed range of companion mass ratio (or mass range) for sub-stellar objects and orbital separation range from $<$ 0.3 AU to $>$ 300 AU.  We utilize results primarily from radial velocity and direct imaging surveys.  Radial velocity survey samples are comprised of nearby main sequence stars MFGK, or evolved A star samples from the field.  Direct imaging survey samples are primarily stars younger than 300 Myr old.  We implicitly assume no significant evolution in the companion mass or orbital distributions over time.  We quote the median value and error bars associated with the 68 \% confidence interval for the published estimates.  Care was taken to only use data which were independent, and carefully consider the errors.  In many cases, specific models are adopted within the original literature in order to derive these estimates (e.g. used to estimate completeness limits) and these are listed in Table 2.  In all cases we kept, the same functional form was adopted for the mass function as well as the orbital distribution.  We fit for the average number of companions per star \footnote{This is defined as the companion frequency $CF = (N_{binaries} + 2N_{triples} + 3N_{quadruples} + ...)/N_{total \ number \ of \ primaries}$.} rather than the probability of having one or more companions.  In four 
    cases, marked in Table 2, the probability of having one or more companions was used.  
    We justify keeping these data based on the small number of systems known with multiple 
    gas giants.  We also tested the robustness of fits for our fiducial model and they did not 
    change when these four data were excluded.  \par

    Host star mass distributions were only available for about 20 \% of our sample.  For this reason, we adopt the mid-range of each sample given as the host mass for that sub-sample.  We also assumed this mid-range mass to convert periods (from 6.25 to 100 days, limited to highest signal--to--noise data) to semi-major axes and to define the q--range of each study.  
     
     
     This resulted in a sample of 51 data points covering companion masses from 0.3 to 85 Jupiter masses, spanning orbital separations from 0.04 to 400 AU.  These data are presented in Table 2.  

\begin{table}[htbp]
\renewcommand{\arraystretch}{1.2}
\centering
\caption{Data base of companion frequencies}
\makebox[\textwidth]{
\scalebox{0.65}{\begin{tabular}{lcccccc}
\hline \hline
$f$ &  $a\,(physical)$ & $q$ & $M_*$ & $m_p$ & Assumed Orbital/Mass  &Reference\\
$\%$ & [AU] & $[\frac{m_p}{M_*}] \times 10^{-3}$ & $M_\odot$ & $M_{\rm Jup}$ & Distribution & \\
\hline
$1^{+2.2}_{-0.3}$ & 75 - 300 & 5 - 10 & 1.04 - 2.1 & 5 - 10 & Linear flat &  \cite{McCarthy_2004}\\
$0.8^{+1.5}_{-0.8}$ & 120 - 1200 & 12 - 75 & 1.04 - 2.1 & 12 - 75 & Linear flat & \cite{McCarthy_2004}\\

$3.2^{+1.7}_{-3.1}$ & 37.8 - 2146.5 & 7.1 - 80 & 0.9-1.7 & 12.57 - 75.42 & Linear flat & \cite{Metchev_2009} \\

$4.3^{+9.1}_{-1.3}\,^a$ & 5 - 320 & 0.95 - 8.91 & 1.5 - 3.0 & 3 - 14 & Linear flat & \cite{Vigan_2012} \\
$2.8^{+6.0}_{-0.9}\,^a$ & 5 - 320 & 4.45 - 47.73 & 1.5 - 3.0 & 14 - 75 & Linear flat & \cite{Vigan_2012} \\

$2^{+3}_{-1}$ & 0.06 - 0.28 & 0.48 - 32.88 & 0.09 - 0.6 & 0.3 - 3.1 & Log-flat  & \cite{Bonfils_2013}  \\


$7.4^{+3.6}_{-2.4}$ & 5 - 320 & 1.43 - 13.36 & 1.0 - 2.0 & 3 - 14 & Linear flat & \cite{Rameau_2013}  \\

$6.5 \pm 3$ & 0 - 27 & 7 - 130 & 0.1 - 0.64 & 1 - 13 & Power-law   & \cite{Montet_2014}\\

$3.9^{+4.8}_{-2.6}$ & 1 - 10 & 20.68 - 715.95 & 0.1 - 0.6 &  13 - 75 & Log-flat& \cite{Bowler_2015}  \\ 
$2.8^{+2.4}_{-1.5}$ & 10 - 100 & 20.68 - 715.95 & 0.1 - 0.6 & 13 - 75 &Log-flat& \cite{Bowler_2015}  \\ 
$\leq 4.1$ & 10 - 100 & 7.96 - 124.10 & 0.1 - 0.6 & 5 - 13 & Log-flat&  \cite{Bowler_2015}  \\ 
$4.4^{+3.2}_{-1.3}\,^a$ & 8 - 400 & 4.79-191.61& 0.01 - 0.8  & 2 - 80 & Linear flat &  \cite{Lannier_2016}\\
$2.1^{+1.95}_{-0.6}\,^a$ & 20 - 300 & 0.32 - 95.46 & 0.75 - 1.5 & 0.5 - 75 & Linear flat & \cite{Vigan_2017} \\

$0.9^{+1.9}_{-0.9}$ & 0.02 - 0.10 & 0.64 - 12.41 & 1.0 - 1.5 & 1 - 13 & Linear flat & \cite{Borgniet_2019} \\
$0.9^{+2.0}_{-0.9}$ & 0.09 - 0.48 & 0.64 - 12.41 & 1.0 - 1.5 & 1 - 13 & Linear flat&  \cite{Borgniet_2019} \\
$4.1^{+3.0}_{-1.2}$ & 0.42 - 2.24 & 0.64 - 12.41 & 1.0 - 1.5 & 1 - 13 & Linear flat&  \cite{Borgniet_2019}\\
$1.4^{+3.1}_{-1.4}$ & 0.02 - 0.12 & 0.38 - 8.27 & 1.5 - 2.5 & 1 - 13 & Linear flat& \cite{Borgniet_2019} \\
$2.0^{+4.5}_{-2.0}$ & 0.10 - 0.57 & 0.38 - 8.27 & 1.5 - 2.5 & 1 - 13 & Linear flat&  \cite{Borgniet_2019} \\
$3.2^{+7.4}_{-3.2}$ & 0.48 - 2.66 & 0.38 - 8.27 & 1.5 - 2.5 & 1 - 13 & Linear flat& \cite{Borgniet_2019} \\
$0.8^{+1.9}_{-0.8}$ & 0.02 - 0.10 & 8.27 - 76.37 & 1.0 - 1.5 & 13 - 80 & Linear flat& \cite{Borgniet_2019} \\
$0.8^{+1.9}_{-0.8}$ & 0.09 - 0.48 & 8.27 - 76.37 & 1.0 - 1.5 & 13 - 80 & Linear flat& \cite{Borgniet_2019} \\
$0.9^{+1.9}_{-0.9}$ & 0.42 - 2.24 & 8.27 - 76.37 & 1.0 - 1.5 & 13 - 80 & Linear flat&  \cite{Borgniet_2019} \\
$1.0^{+2.2}_{-1.0}$ & 0.02 - 0.12 & 4.96 - 50.91 & 1.5 - 2.5 & 13 - 80 & Linear flat&  \cite{Borgniet_2019} \\
$1.1^{+2.3}_{-1.1}$ & 0.10 - 0.57 & 4.96 - 50.91 & 1.5 - 2.5 & 13 - 80 & Linear flat&  \cite{Borgniet_2019} \\
$1.3^{+2.8}_{-1.3}$ & 0.48 - 2.66 & 4.96 - 50.91 & 1.5 - 2.5 &13 - 80 & Linear flat& \cite{Borgniet_2019} \\

$7.8^{+9.1}_{-3.3}$ & 0.27 - 2.87 & 0.48 - 8.38 & 1.06 - 1.59 & 0.8 - 9.3 & Linear flat &  \cite{Wittenmyer_2019} \\

$0.5^{+1.4}_{-0.2}$ & 0.02 - 0.04 & 0.17 - 27.58 & 0.45 - 1.72 & 0.3-13 & Linear flat &\cite{Wittenmyer_2019} \\
$1^{+1.6}_{-0.5}$ & 0.04 - 0.09 & 0.17 - 27.58 &  0.45 - 1.72 & 0.3-13 & Linear flat&  \cite{Wittenmyer_2019} \\
$1^{+1.6}_{-0.5}$ & 0.09 - 0.19 & 0.17 - 27.58 &  0.45 - 1.72 & 0.3-13 & Linear flat& \cite{Wittenmyer_2019} \\
$\leq 1.2$ & 0.19-0.42 & 0.17 - 27.58 &  0.45 - 1.72 & 0.3-13 & Linear flat& \cite{Wittenmyer_2019} \\
$1.7^{+1.9}_{-0.7}$ & 0.42 - 0.88 & 0.17 - 27.58 &  0.45 - 1.72 & 0.3-13 & Linear flat&\cite{Wittenmyer_2019} \\
$8^{+2.7}_{-2.2}$ & 0.88 - 1.96 & 0.17 - 27.58 &  0.45 - 1.72 & 0.3-13 & Linear flat& \cite{Wittenmyer_2019} \\
$5.3^{+2.8}_{-1.5}$ & 1.96 - 4.07 & 0.17 - 27.58 &  0.45 - 1.72 & 0.3-13 & Linear flat& \cite{Wittenmyer_2019} \\
$6.9^{+3.2}_{-2.1}$ & 4.07 - 9.08 & 0.17 - 27.58 &  0.45 - 1.72 & 0.3-13 & Linear flat& \cite{Wittenmyer_2019} \\


$3.5^{+1.9}_{-1.4}$ & 10 - 100 & 0.95 - 62.03 & 0.2 - 5.0 & 5 - 13 & Power-law & \cite{Nielsen_2019} \\ 
$0.8^{+0.8}_{-0.5}$ & 10 - 100 & 2.48 - 381.71 & 0.2 - 5.0 & 13 - 80 & Power-law & \cite{Nielsen_2019} \\ 

$14.1^{+2.0}_{-1.8}$ & 2 - 8 & 0.06 - 63.64 & 0.3 - 1.5 & 0.1 - 20 & Log-flat & \cite{2021ApJS..255....8R}\\
$8.9^{+3.0}_{-1.4}$ & 8 - 32 & 0.06 - 63.64 & 0.3 - 1.5 & 0.1 - 20 & Log-flat & \cite{2021ApJS..255....8R} \\

$3^{+2}_{-1}$ & 0.1 - 1 & 0.07 - 18.72 & 0.51 - 1.39 & 0.1 - 10 & Log-flat &  \cite{2021AJ....161..134H} \\ 
$18 \pm 4$ & 1 - 10 & 0.07 - 18.72 & 0.51 - 1.39 & 0.1 - 10 & Log-flat & \cite{2021AJ....161..134H}  \\

$4^{+3}_{-2}$ & 0.04 - 0.37 & 0.46 - 33.74 & 0.09 - 0.65 & 0.32 - 3.2 &  Power-law  & \cite{Sabotta_2021} \\ 
$5^{+4}_{-3}$ & 0.29 - 1.70 & 0.46 - 33.74 & 0.09 - 0.65 & 0.32 - 3.2 & Power-law & \cite{Sabotta_2021} \\

$23.0_{-9.7}^{+13.5}$ & 5 - 300 & 333 - 2500 & 1.42-2.09 & 1 - 75 & Linear Flat &  \cite{Vigan_2021}\\
$5.8_{-2.8}^{+4.7}$ & 5 - 300 & 2 - 250 & 0.89-1.45 & 1 - 75 & Linear Flat & \cite{Vigan_2021}\\
$12.6_{-7.1}^{+12.9}$ & 5 - 300 & 0.263 - 395 & 0.3-0.5 & 1 - 75 & Linear Flat & \cite{Vigan_2021}\\

$10.7_{-1.6}^{+2.2}$ & 0.1212 - 1.258 & 0.34 - 21.36 & 1.09 - 2.27 & 0.8 - 24.4 & Log-normal & \cite{Wolthoff_2022}\\

$3.4^{+5.4}_{-2.5}$ & 81 - 108 & 2 - 10 & 1.5 - 2.5 & 5 - 15 & Linear Flat & \cite{Wagner_2022} \\

$19.2\pm 2.8$ & 1 - 10 & 0.67 - 20 & 1.0 - 1.5 & 0.3 - 10 & Power-law & \cite{Zhu_2022}\\

$0.7^{+1.6}_{-0.2}$ & 0.00785 - 0.987 & 1 - 13 & 0.76 - 1.20 & 1 - 13 & Linear flat &  \cite{Grandjean_2023}\\
$0.6^{+1.4}_{-0.2}$ & 0.00785 - 0.987 & 21.7 - 133 & 0.52 - 0.68 &  13 - 80 & Linear flat & \cite{Grandjean_2023}\\
$1.4^{+3.2}_{-0.4}$ & 0.00785 - 0.987 & 8.67 - 53.3 & 1.19 - 1.81 & 13 - 80 & Linear flat &  \cite{Grandjean_2023}\\

\hline
\label{tab:database}
\end{tabular}}
}
{\small $\,$ $^a$ frequency of host having one or more companions, not the mean number of companions per star.}
\end{table}

    \subsection{The Fitting Procedure:}
    
    For each model, we can assume values for the parameters, and integrate the functions in equation 1 to give a frequency of companions over every range of mass and separation realized in the database.  These estimates are then compared to the data.  We explore a range of parameters for each variable and try to find the best ensemble fit, using a variety of 
    metrics, for each model making use of this likelihood function:

\begin{equation}
\log \mathcal{L}(\theta) = -\sum_i \log \mathcal{L}_i(\theta) = -\sum_i
\begin{cases}
\displaystyle \frac{(f_{{\rm obs}, i} - f_{{\rm model}, i}(\theta))^2}{2 \sigma_i^2}, & \text{Gaussian} \\
\displaystyle -\log\left( \frac{1}{2} \cdot \mathrm{erfc}\left( \frac{f_{{\rm model}, i}(\theta) - f_{{\rm obs}, i}}{\sigma_i} \right) \right), & \text{Upper limit} \\
\displaystyle -\log\left( \frac{P(f_{{\rm model}, i}(\theta) \mid s_i, \mathrm{offset}_i, \mathrm{scale}_i)}{P(f_{\mathrm{peak}, i} \mid s_i, \mathrm{offset}_i, \mathrm{scale}_i)} \right), & \text{Log-normal}
\end{cases}
\label{eq:likelihood}
\end{equation}

Here, $\theta$ represents the set of model parameters, and for each data point $i$, $f_{{\rm model}, i}(\theta)$ is the predicted companion frequency, $f_{{\rm obs}, i}$ is the observed frequency, and $\sigma_i$ is the associated uncertainty. For data with Gaussian errors, we assume normally distributed uncertainties. For upper limits, the likelihood is computed using the complementary error function $\mathrm{erfc}$ to evaluate the probability that the model prediction does not exceed the observational threshold. For log-normal errors, we evaluate the log-likelihood using the log-normal probability density function defined by the shape parameter $s$, offset, and scale. The PDF is given by 
\[
P(f) = \frac{1}{(f - \mathrm{offset}) \cdot s \cdot \sqrt{2\pi}} \exp\left( -\frac{(\log(f - \mathrm{offset}) - \log(\mathrm{scale}))^2}{2s^2} \right),
\]
and is normalized by its mode, $f_{\mathrm{peak}} = \exp(\log(\mathrm{scale}) - s^2) + \mathrm{offset}$. This unified likelihood function allows consistent fitting of the model across heterogeneous data types.

    
    We used the MultiNest algorithm \citep{2009MNRAS.398.1601F, Feroz_2009, 2019OJAp....2E..10F} to explore the posterior distributions of our model parameters. We used the PyMultiNest Python wrapper \citep{Buchner2016} to interface with the sampler. Using the nested sampling technique, MultiNest provides robust estimates of the marginalized posterior distributions, although the exact location of the global maximum likelihood estimate (MLE) is determined by the highest-likelihood point found during the sampling.
    We assess goodness of fit for the best fitting model using a variety of metrics.  We compute the reduced chi--square, indicating the probability that these data were drawn from the model in a frequentist sense.  We also calculate the Bayesian Information Criteria (BIC): 
    
    $$\mathrm{BIC} = k\ln(n) - 2\ln(\widehat{L})$$
    
    where  k is the number of free parameters in the model, n is the number of independent data points, and L is the likelihood of the model built from MLE parameter values.  We prefer to use the f--test, comparing the likelihood of a model with $N+1$ parameters to a comparable nested model with only $N$ parameters, whenever possible.  The f--test provides a more robust model selection than the BIC.  MultiNest also provides the Bayesian evidence, enabling 
    users to assess relative probabilities of different models. We compare these metrics for 
    model selection as discussed below. 

    
\section{Results}

First we assess the fit of our fiducial model to the data as well as other functional forms concluding that our fiducial model is the best.  The fiducial model has six parameters: $A_{p}$, $\sigma_{p}$, $\mu_{p}$, $\alpha$, $A_{bd}$, and $\beta$ (four for the planet term and two for the brown dwarf term).  The planet companion masses have an upper limit at 10 \% the mass of the star, and the brown dwarf companion masses have a lower limit due to the opacity limit for fragmentation.  We compare this model to another six parameter model, replacing $\sigma_{pl}$ and $\mu_{pl}$ with a power--law orbital distribution with a cutoff radius (\cite{2016PASP..128j2001B}).  We then use results from the fiducial model to explore: a) whether the planet mass function depends on orbital separation; and b) whether we can discern any evidence that the orbital separation distribution for planets depends on host star mass.

\subsection{Best Fits and Model Selection}

Starting with the fiducial model described above, we fixed the opacity limit for fragmentation to 3, 10, and 30 Jupiter mass for three separate fits respectively.  For the MLE in each case, the 
reduced chi--squared was 1.01, 0.98, and 0.98, with BIC, and Bayesian evidence tracking the 
reduced chi--squared:  the fits are of equal quality and all six parameters from each fit are 
consistent with each other.  We tried 
to fit for the cutoff mass in a seven parameter model resulting in a comparable fit with a preferred cutoff mass of 15 Jupiter masses (though the parameter is essentially unconstrained based on the flat PDF).  Based on the f--test, the addition of this parameter is not justified compared to the fixed cutoff. We also tried to fit a model with six parameters without a cutoff mass for brown dwarf companions.  This model also gave a similar fit.  We checked for sensitivity of the result to modest changes of the priors as well as different initial conditions, and the 
fitting was found to be robust.  We provisionally adopt the six-parameter model fit with a brown dwarf cutoff at 3 Jupiter masses.  We adopt the 3 Jupiter mass cutoff based on results from young 
star clusters suggesting an opacity limit for fragmentation near this limit 
(\cite{2025ApJ...981L..34D}; cf. Adams et al. accepted) though other results quoted below are independent of this choice.  

    \begin{figure*}[h]
      \centering
      {\includegraphics[width=\textwidth]{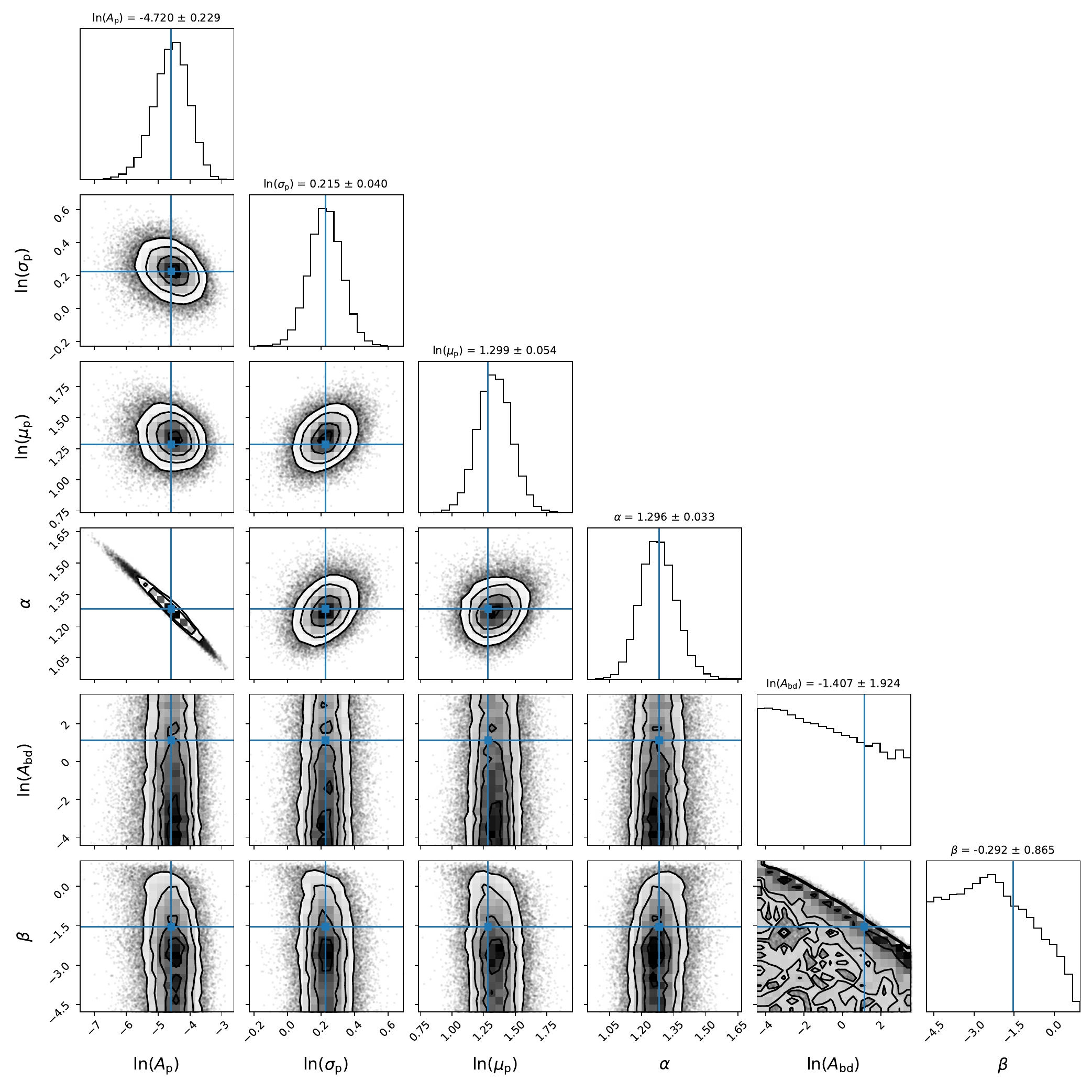}}
      \caption{Corner plot of the posterior PDFs for the six parameters of the fiducial model, generated from the MultiNest posterior samples. The maximum likelihood values  are quoted at the top of each column, which we adopt as the best fit model, as outlined in Table 3. The blue lines and squares show the median values for each parameter.}
      \label{Fig. 1}
    \end{figure*}

    \begin{figure*}[h]
      \centering
      {\includegraphics[width=.8\textwidth]{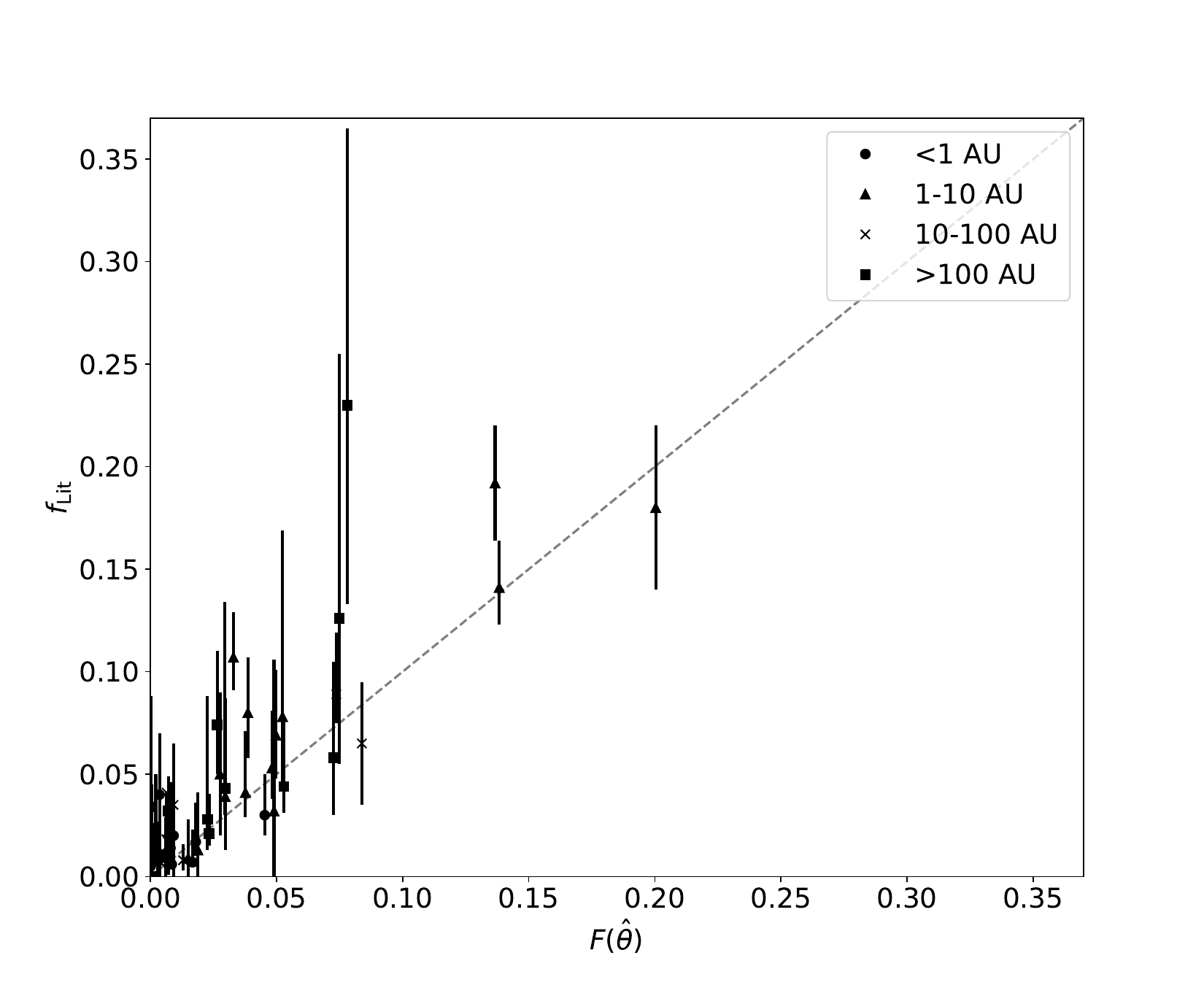}}
      \caption{Comparison between Model \#1 frequency predictions from the MLE parameters and the data from the literature. The diagonal line shows the 1:1 relationship. Data points with orbital separations below 1 AU are shown by circle symbols, and data points with separations that stretch beyond 100 AU are represented by squares.}
      \label{Fig. 2}
    \end{figure*}


    Figure 1 shows the corner plot for this fiducial model and the resulting probability density functions (hereafter PDF).  In Figure 2 we show the predictions of the fiducial model based on the MLE values from Figure 1 (on the x-axis) compared to all the data (y-axis) with uncertainties from Table 1.  The fits were done in natural log space as summarized in Table 3.  The power-law slope for the mass function of planets $\alpha$ is consistent with \cite{2008PASP..120..531C}.   This is not surprising, as the result is consistent with a number of radial velocity studies, many of which are used in this 
    work.  However quantitative constraints confirming this are lacking in the literature.  Regarding the shape of the gas giant planet orbital distribution, we find a characteristic separation of 3.8 AU.  This compares well with studies based on sub-sets of these data (\cite{2018A&A...612L...3M} and \citet{2021ApJS..255...14F}; cf \cite{2019ApJ...874...81F}). The width of the best fit log--normal is consistent within the errors to that reported in \cite{2018A&A...612L...3M} for 
    M dwarfs.  The data for M, FGK, and A stars can all be fitted with a single normalization for gas giant planets and a single normalization for brown dwarf companions, when considered over appropriate ranges of q (\cite{2022A&A...657A..48S}).  
    
    The distributions of $A_{bd}$ and $\beta$ for the brown dwarf term are consistent with those fitted for stellar companions (Table 1), but are not well constrained.  We tried fixing either $A_{bd}$ or $\beta$, but the fits were poor ($\chi_{\nu}^2 > 2)$.  We also explored various orbital distributions for the brown dwarf companions including log-flat with a flat brown dwarf companion mass function, which gave a terrible fit.  Finally, we tried a nine parameter model, our six parameters from the fiducial model, plus a brown dwarf cutoff, as well as two additional parameters for a log--normal brown dwarf orbital distribution ($\mu_{bd}$ and $\sigma_{bd}$). While this fit was good ($\chi_{\nu}^2$ = 1.197), the additional three parameters were largely unconstrained.  Further, the f--test suggested that the fit was not better enough to warrant the additional parameters.  In summary, we could not get a fit where $A_{bd}$ and $\beta$ were well constrained, though the best fit is consistent with simple extrapolations of the stellar companion distributions.  This could be interpreted as the so--called brown dwarf desert being simply due to the limited range of q below 0.1 given a flat companion mass ratio distribution.  

We adopt the MLE from the six parameter fit shown in Figure 1 and Table 3 as our fiducial model. 
The errors for the MLE are estimated fitting a gaussian to the local curvature in likelihood 
space near the MLE. 
Recall we assume an upper mass cutoff for planets at 10 \% the mass of the star, and
a lower mass cutoff for brown dwarfs at 3 Jupiter mass.  This limit is near the 
opacity limit for fragmentation (Adams et al. accepted) and was recently found to 
be the turnover mass in the very young cluster NGC 2024 (De Furio et al. 2025). Yields from this model can be estimated with user defined inputs using the online tool found at this location:
https://meyer2025demographics.streamlit.app/.

We now compare this fit to one where the planet orbital distribution is assumed to 
be a the power--law in orbital separation with outer cut--off.  This model has a reduced chi--squared of 1.36 corresponding to  probability of 0.05 (for 51 - 6 = 45 degrees of freedom) 
that the data came from this model, 
compared to a p--value of 0.46 for our fiducial model.  While \cite{2018A&A...612L...3M} were unable to distinguish between the log--normal fit and the power--law with cut--off (cf. \citep{2016ApJ...817..104R}, we strongly favor the log--normal.   

To check whether functional fits in the mass ratio are superior or not to functional 
fits in companion mass with a stellar mass dependence in the normalization coefficient, 
we did the following.  We compared the fiducial six parameter fit to a seven parameter 
fit where we fit the mass function for planets in units of planet mass, but with 
a normalization factor that depends on host star mass to the $\gamma$.  The seven 
parameter model has a $\chi_{\nu}^2$ = 1.6 indicating a worse fit as well as the BIC. 
Further, $\gamma = 0.89^{+0.05}_{-0.18}$: if $\gamma = \alpha$ (found to be 1.30 $\pm$ 0.03 in our fiducial model) this seven parameter model reduces identically to 
the fiducial model. Finally, the f--test between these two nested models fitted to the same data indicates the seventh parameter is not justified.  Because of the results of the f--test, and the consistency between the values of $\gamma$ and $\alpha$, we prefer to adopt the 
simpler 6--parameter model.  However we caution this is a choice, rather than 
a conclusion strongly indicated from the fitting. 

  \begin{table}[h]
    \begin{threeparttable}
        \caption{Maximum Likelihood Estimate with Gaussian Error Estimates for Fiducial Model}
        \label{table:2}
        \centering
        \begin{tabular}{c c c c c c}
            \hline\hline
            $\ln A_{\rm p}$ & $\ln \sigma_{\rm p}$ & $\mu_{\rm p}$ & $\alpha$ & $\ln A_{\rm bd}$ & $\beta$ \\
            \hline
            $-4.72 \pm 0.229$ & 
            $0.215 \pm 0.04$ & 
            $1.30 \pm 0.05$ & 
            $1.30 \pm 0.03$ & 
            $-1.41 \pm 1.92$ & 
            $-0.29 \pm 0.87$ \\
            \hline
        \end{tabular}
        \begin{tablenotes}
            \small
            \item The quoted values are the MLE with errors estimated fitting a gaussian to the local curvature of the likelihood space.  For reference $A_{\rm p} = 0.009$, $ \sigma_{\rm p} = 0.5385$ in $log_{10}$, $\mu_{\rm p} = 0.5646$ in $log_{10}$, $A_{\rm bd} = 0.245$, and $\alpha$ and $\beta$ are the same values.  Because the model was fitted in the 
            natural log, using these values for the normalization coefficients 
            in log--10 space requires a factor of log(e) in the deminator of the function.
        \end{tablenotes}
    \end{threeparttable}
\end{table}

  \subsection{Exploring Dependencies with the Best Fit Model}

To search for evidence that the planet mass function depends on orbital separation, we split the database from Table 2 into two equal halves (0.03--1.33 AU versus 1.57--1090 AU) as a function of orbital separation (26 points versus 25).  We fix five parameters of the fiducial model, all except the power--law of the planet mass function.  Then we separately fit both the inner and the outer distributions.  We get a power--law exponent of $1.29 \pm 0.03$ for the inner distribution versus $1.28 \pm 0.01$ for the outer distribution where the difference is not significant.  This result is consistent with our assumption that the planet mass function does not depend on orbital separation down to 0.3 Jupiter masses, as found by \citep{2021ApJS..255...14F} down to 30 Earth masses. 

The peak of the orbital distribution that best fits the data is near the iceline expected for these host stars.  To explore whether there is a difference in the peak of the orbital separation distribution as a function of host star mass, we did the following.  We fixed five model parameters, this time fitting only the peak of the orbital distribution for two subsamples: the M dwarf host star data from Table 2 (nine points) versus the data from A star hosts (16 points). Both fits were reasonable, with reduced chi--squares of about 1.1.  For the M dwarfs the best fit peak is $\mu_p = 0.55^{+0.08}_{-0.04}$ while for the A stars the best fit peak is $\mu_p = 1.52 \pm 0.04$ with essentially 
zero overlap between the two PDFs.  For the M dwarfs this peak corresponds to 1.7 AU while for the A stars the peak is 4.7 AU where the fit for M--FGK--A stars yields 3.8 AU.  To explore this apparent difference further, we calculated the M dwarf and A star frequencies, 
having subtracted the model predicted brown dwarf 
frequencies for both, for a common range of q assuming the fiducial model planet mass function. 
We calculated the L2 norm between the two normalized PDFs yielding a value of 1.62.  Employing a Kolmogorov--Smirnov (KS) statistic, we estimate the likelihood that the two distributions were drawn from the same parent population to be $< 10^{-5}$, supporting the idea that the distributions are distinct. 
  
  \subsection{Companion Mass Ratio Distributions}
  
    Combining our fitted planetary and brown dwarf companion populations, we can create companion mass ratio distributions as a function of orbital separation and host star mass.  We plot these results in Figures 3--4--5--6.  There are two central points that emerge from these plots.  First, there is often a local minimum in the companion mass ratio distribution between $ 0.001 < q < 0.1$.  Second, because the peak (and width) of the planet orbital distributions compared to the brown dwarf companion orbital distributions vary as a function of host star mass, the relative contribution of planets and brown dwarf companions varies.  Because of this, the local minimum shifts as a function of orbital separation and host star mass.  \par
    
    \begin{figure*}[h]
      \centering
      \subfloat[]{\includegraphics[width=0.3\textwidth]{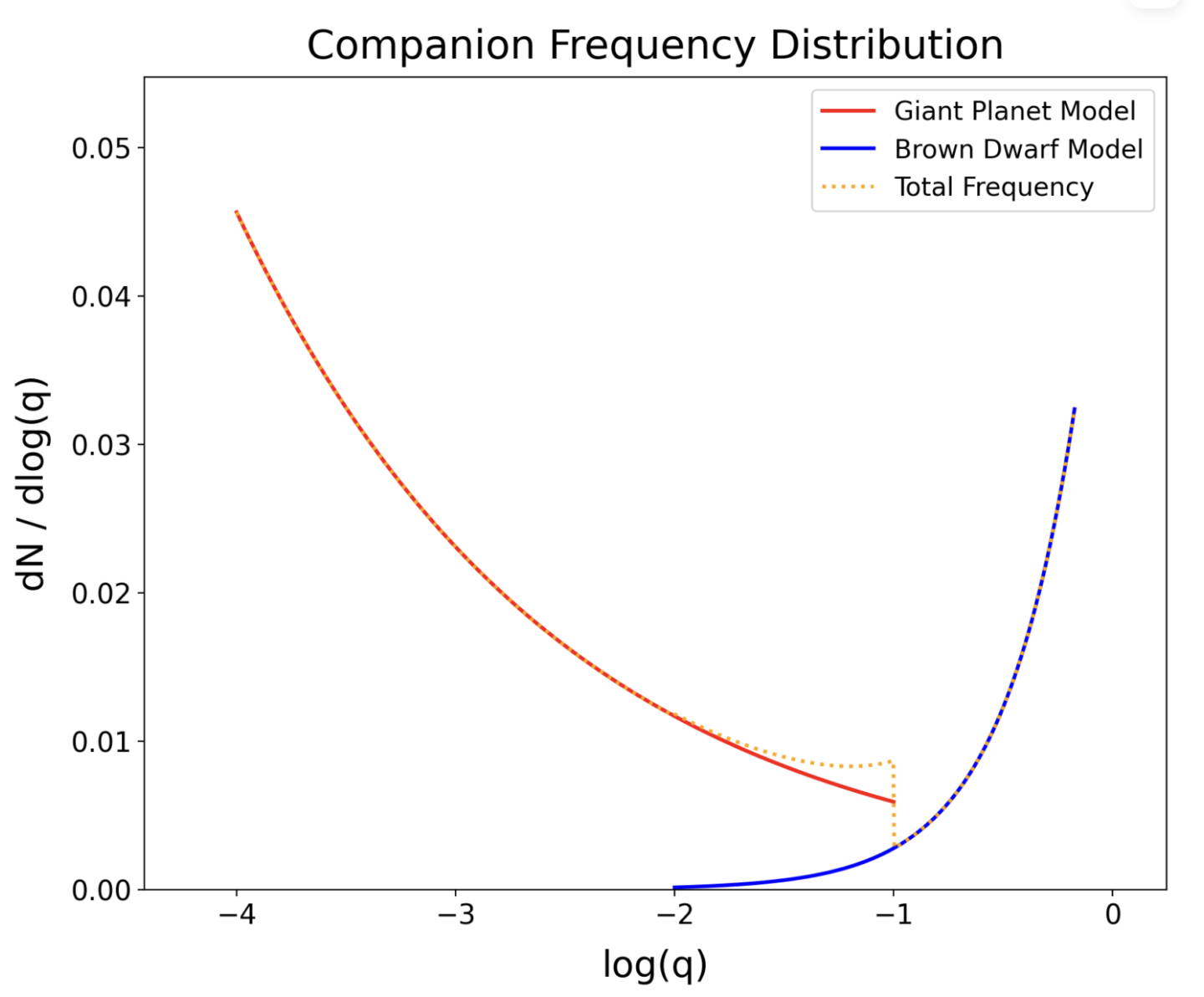} \label{fig:f1}}
      \hspace{0.5em}
      \subfloat[]{\includegraphics[width=0.3\textwidth]{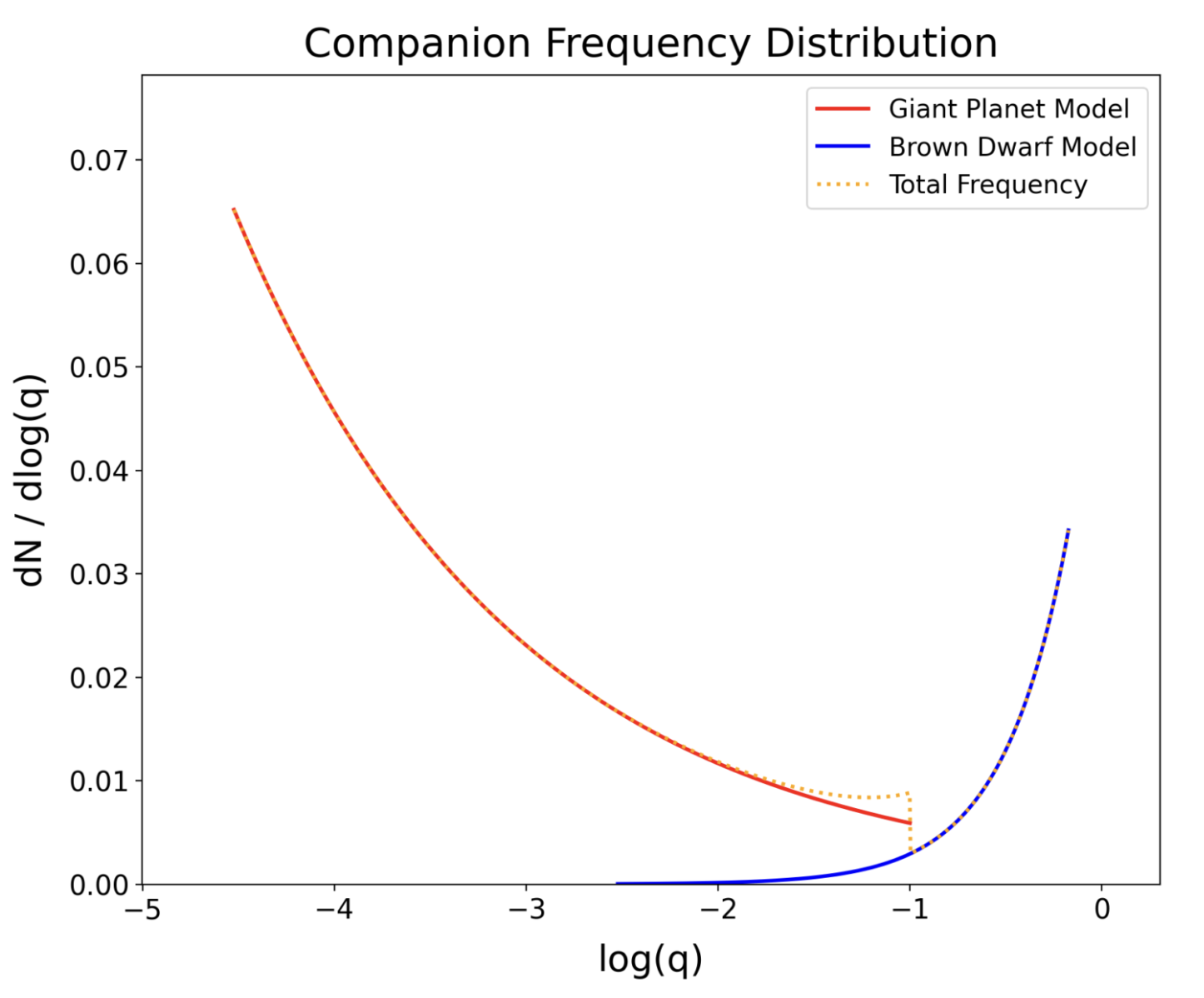} \label{fig:f2}}
      \hspace{0.5em}
      \subfloat[]{\includegraphics[width=0.3\textwidth]{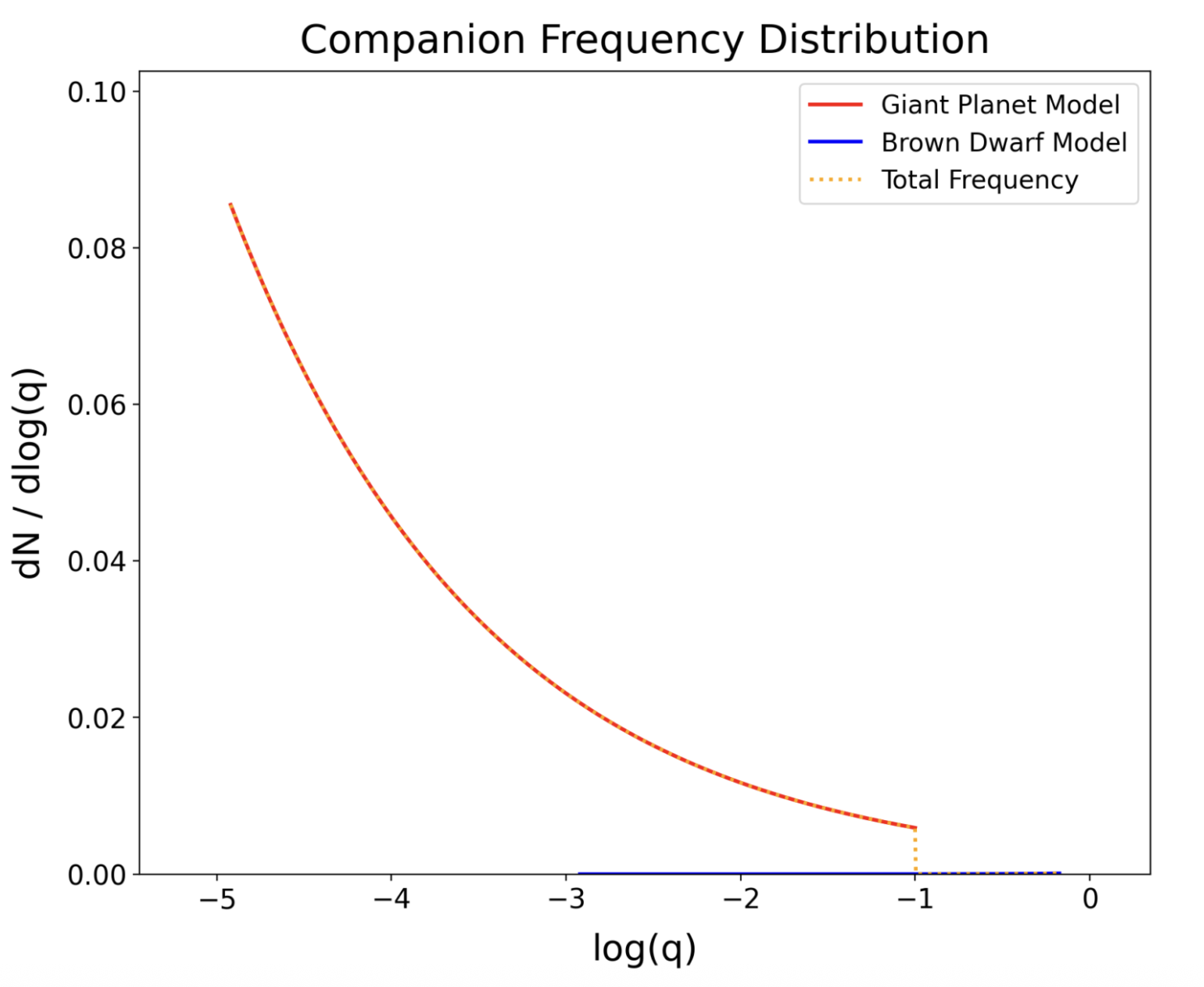} \label{fig:f3}}
      \caption{Companion mass ratio distribution from $-5 < log(q) < -0.2$ for M dwarf (left), FGK stars (middle), and A stars (right) for orbital separations from 0.1-1 AU.  For all panels, brown dwarf binary distributions are in blue, the planet distributions are in red, and orange is the sum.}
      \label{Fig. 3}
    \end{figure*}
    
    \begin{figure*}[h]
      \centering
      \subfloat[]{\includegraphics[width=0.3\textwidth]{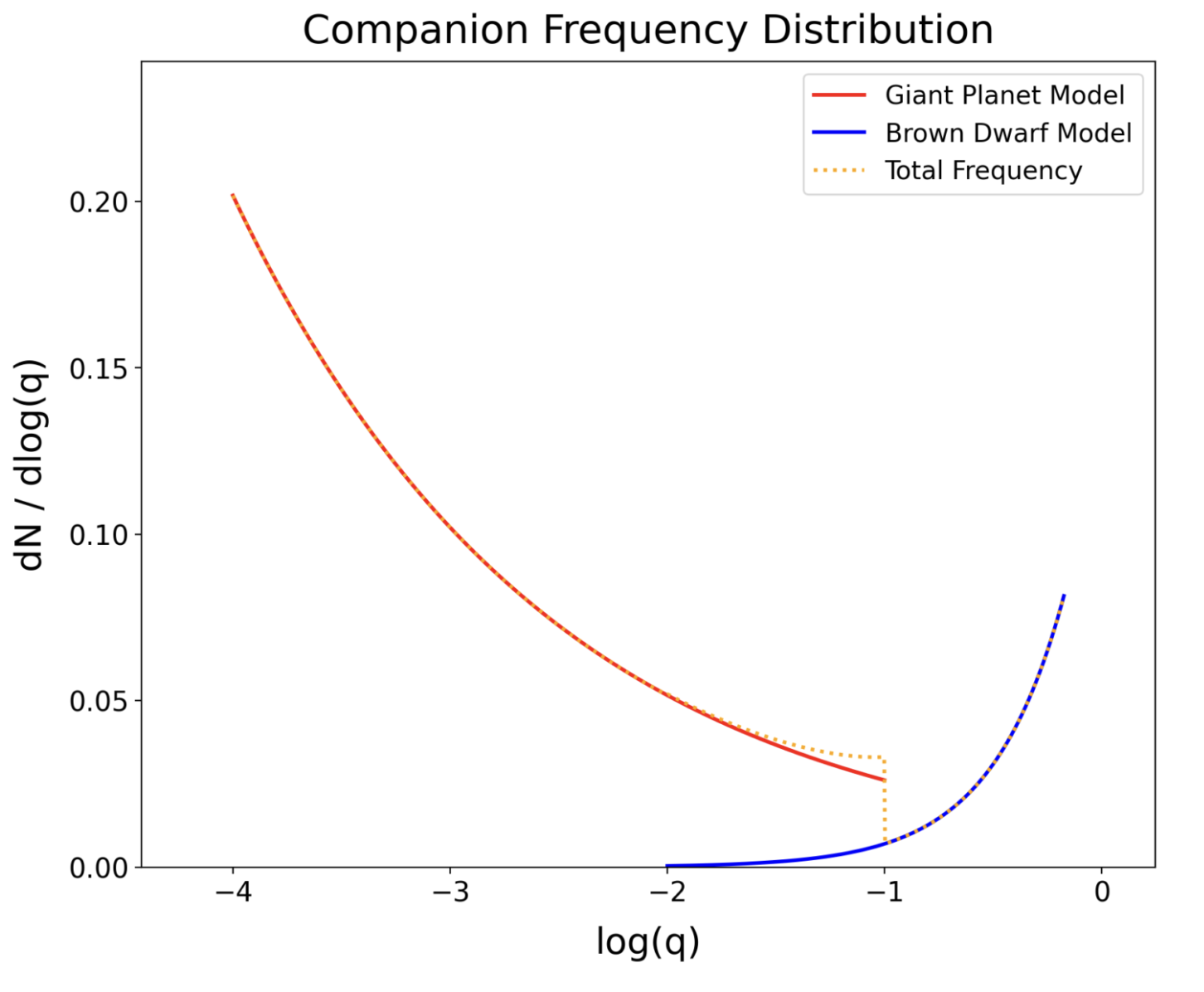} \label{fig:f4}}
      \hspace{0.5em}
      \subfloat[]{\includegraphics[width=0.3\textwidth]{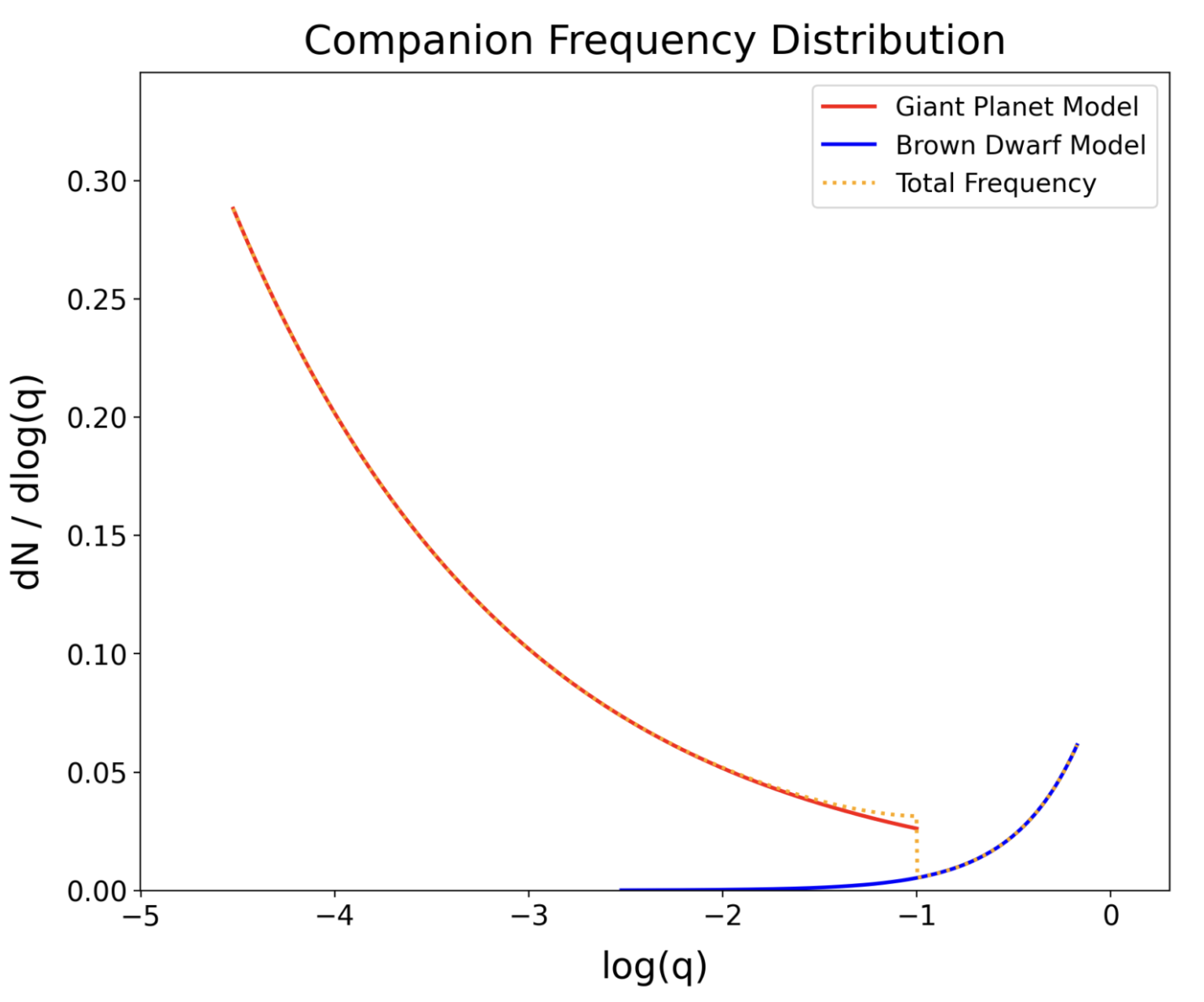} \label{fig:f5}}
      \hspace{0.5em}
      \subfloat[]{\includegraphics[width=0.3\textwidth]{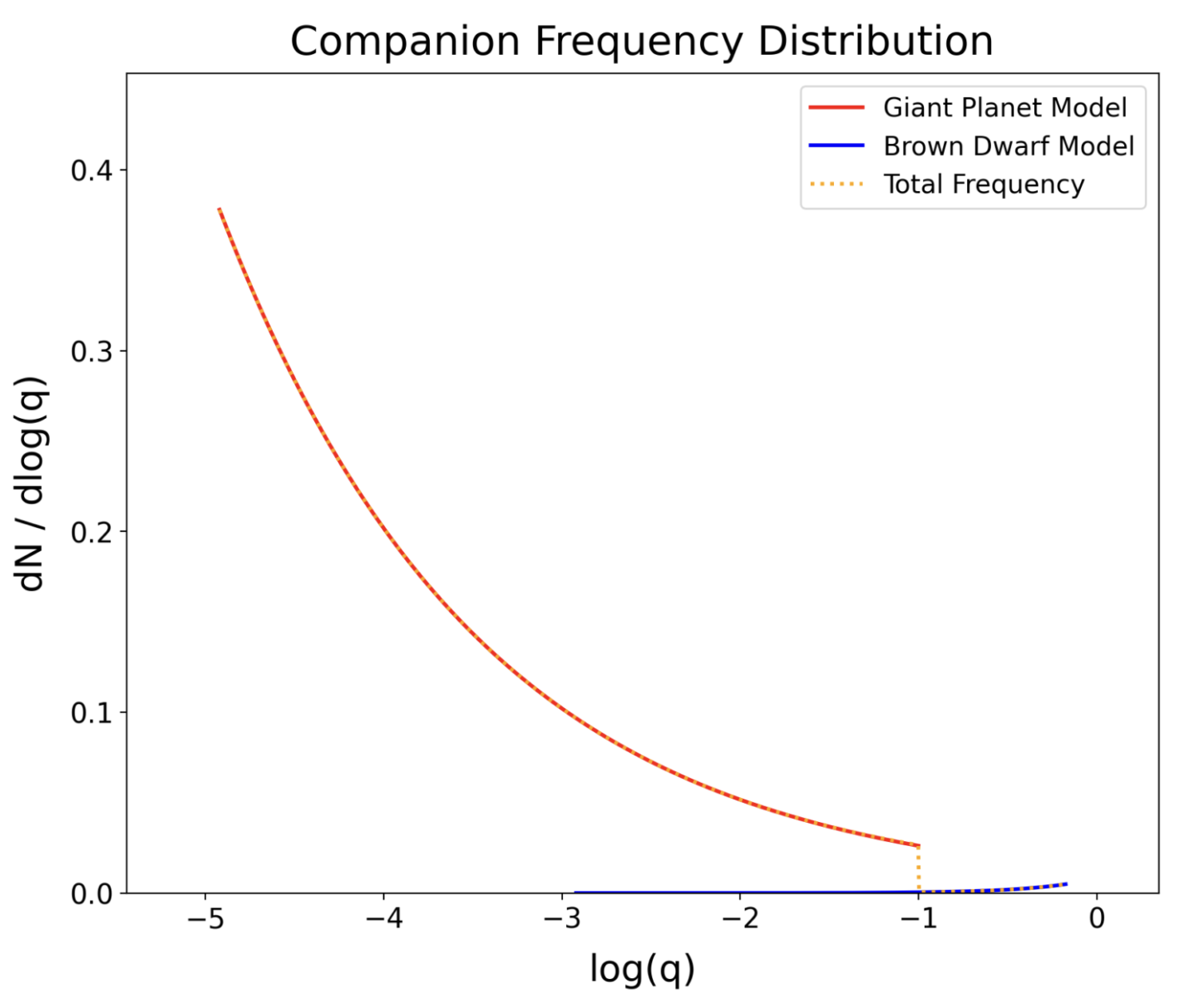} \label{fig:f6}}
      \caption{Companion mass ratio distribution from $-5 < log(q) < -0.2$ for M dwarf (left), FGK stars (middle), and A stars (right) for orbital separations from 1-10 AU.  For all panels, brown dwarf binary distributions are in blue, the planet distributions are in red, and orange is the sum.}
      \label{Fig. 4}
    \end{figure*}
    
   \begin{figure*}[h]
      \centering
      \subfloat[]{\includegraphics[width=0.3\textwidth]{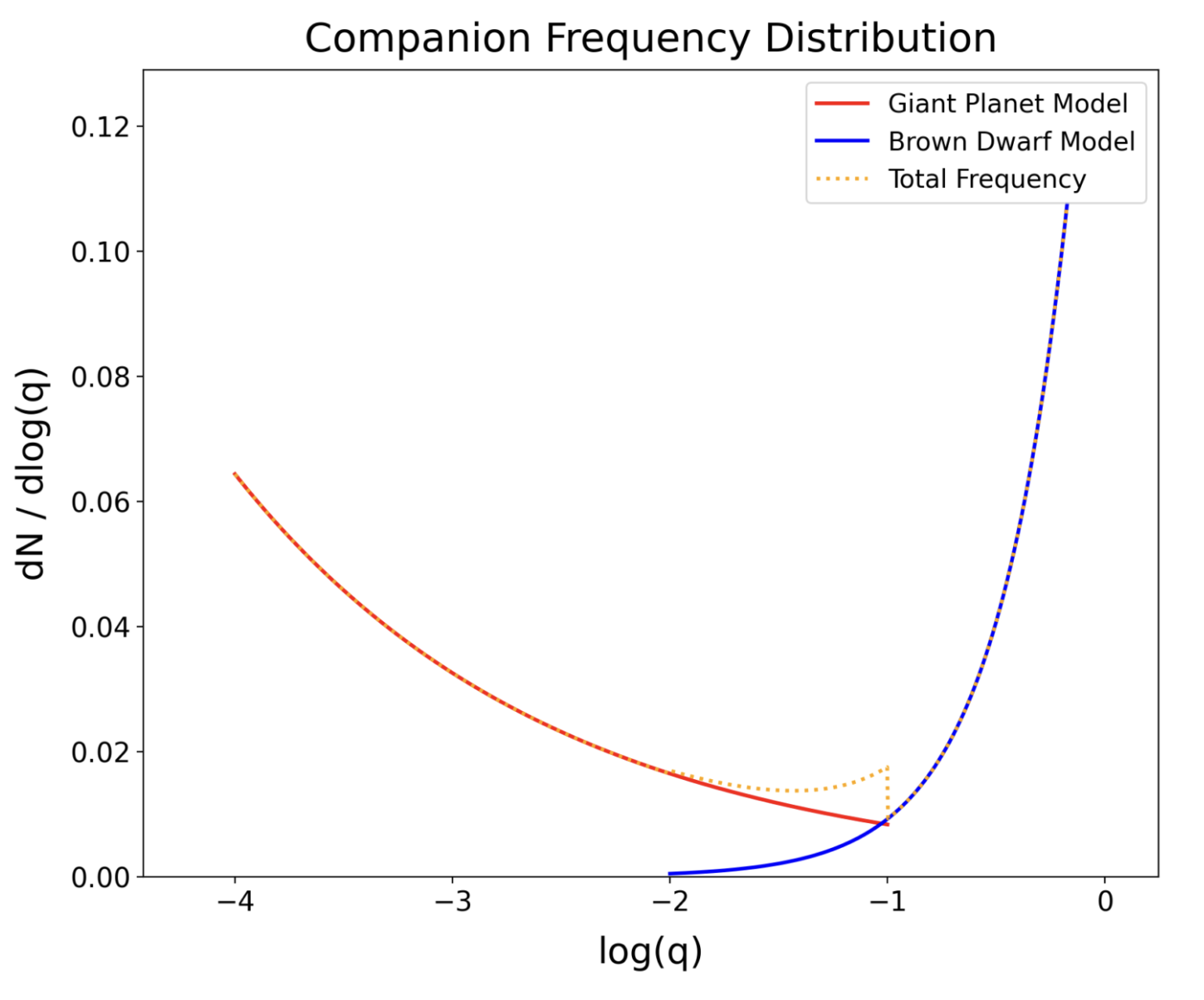} \label{fig:f7}}
      \hspace{0.5em}
      \subfloat[]{\includegraphics[width=0.3\textwidth]{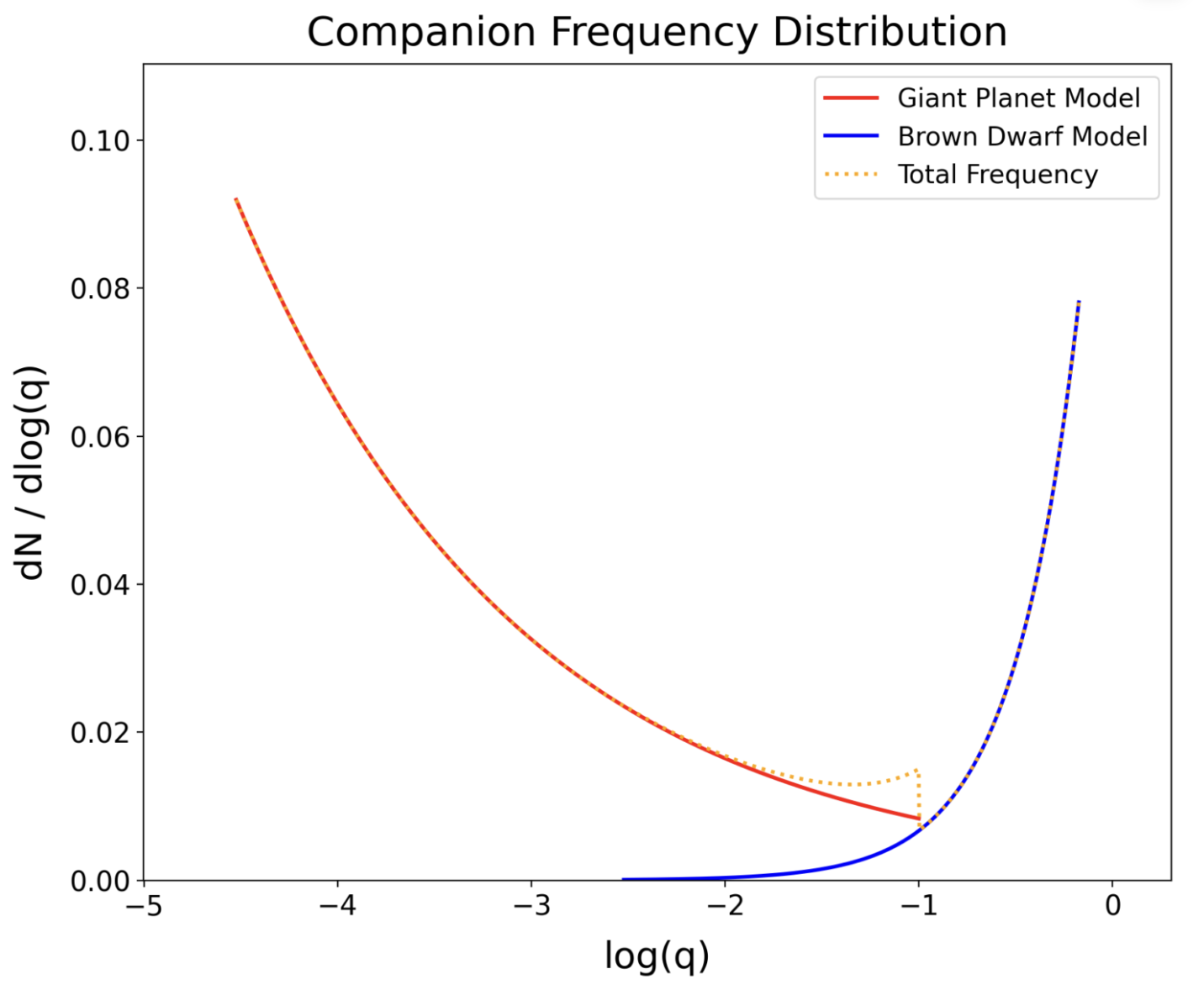} \label{fig:f8}}
      \hspace{0.5em}
      \subfloat[]{\includegraphics[width=0.3\textwidth]{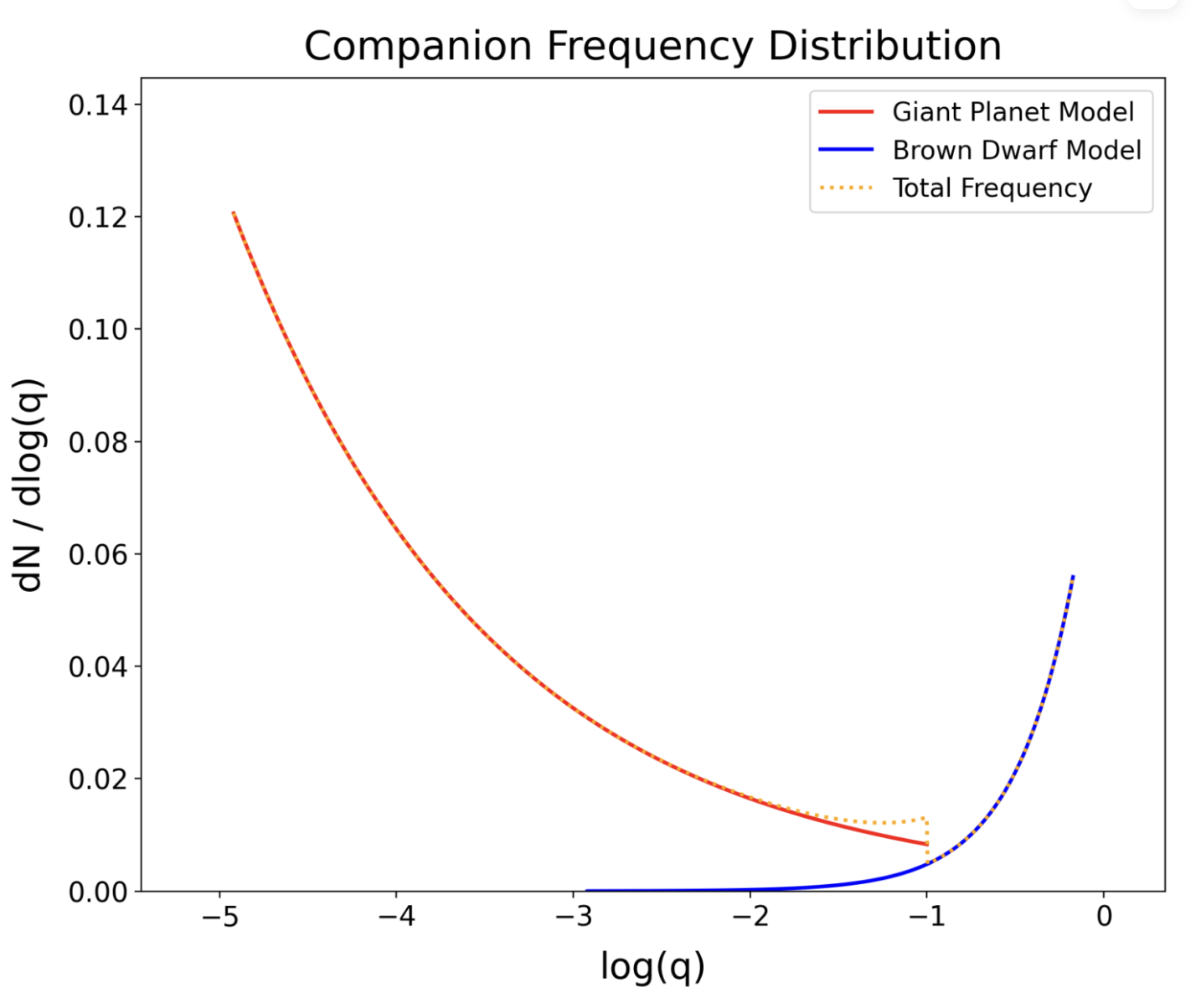} \label{fig:f9}}
      \caption{Companion mass ratio distribution from $-5 < log(q) < -0.2$ for M dwarf (left), FGK stars (middle), and A stars (right) for orbital separations from 10-100 AU.  For all panels, brown dwarf binary distributions are in blue, the planet distributions are in red, and orange is the sum.}
      \label{Fig. 5}
    \end{figure*}
    
    \begin{figure*}[h]
      \centering
      \subfloat[]{\includegraphics[width=0.3\textwidth]{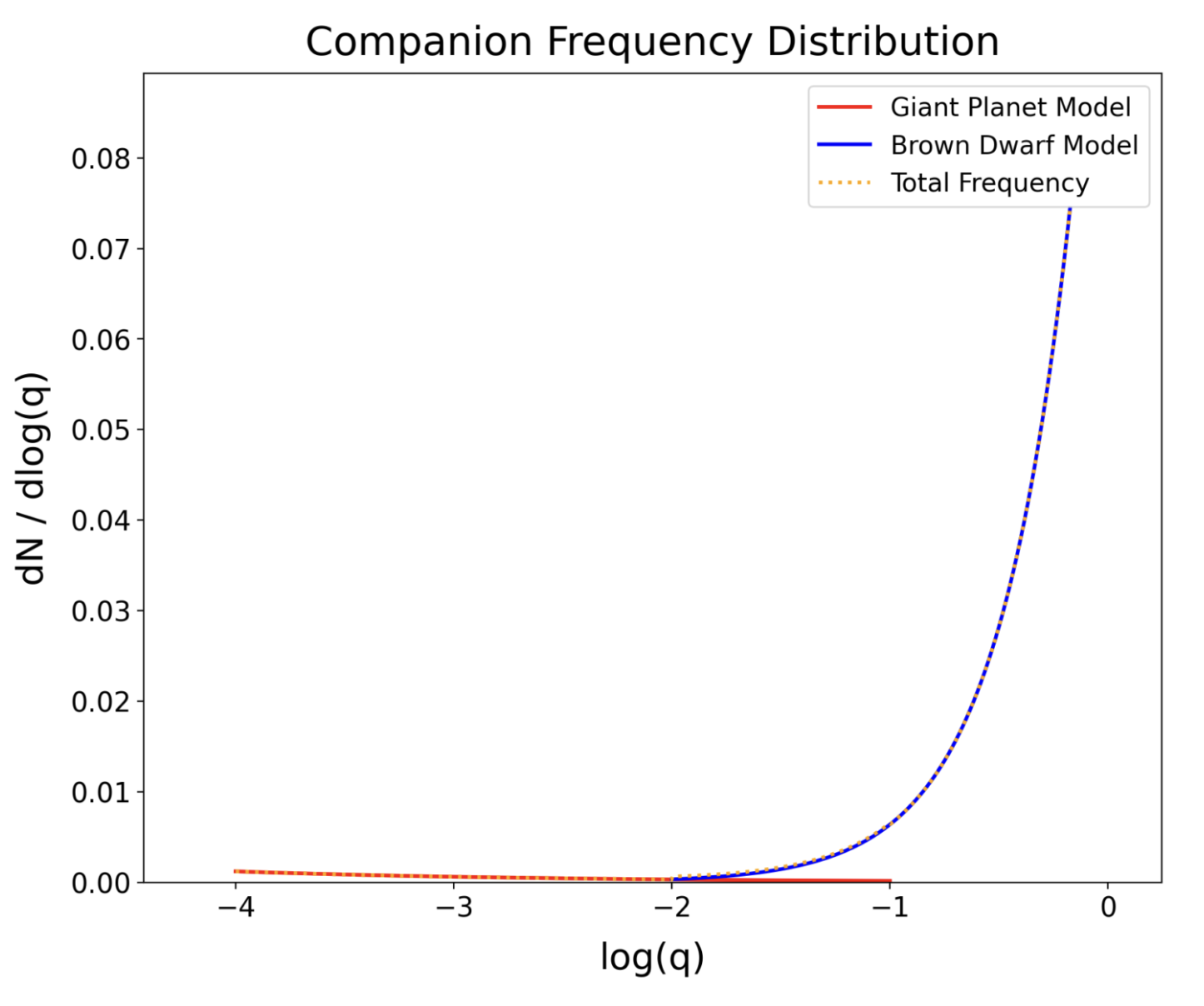} \label{fig:f10}}
      \hspace{0.5em}
      \subfloat[]{\includegraphics[width=0.3\textwidth]{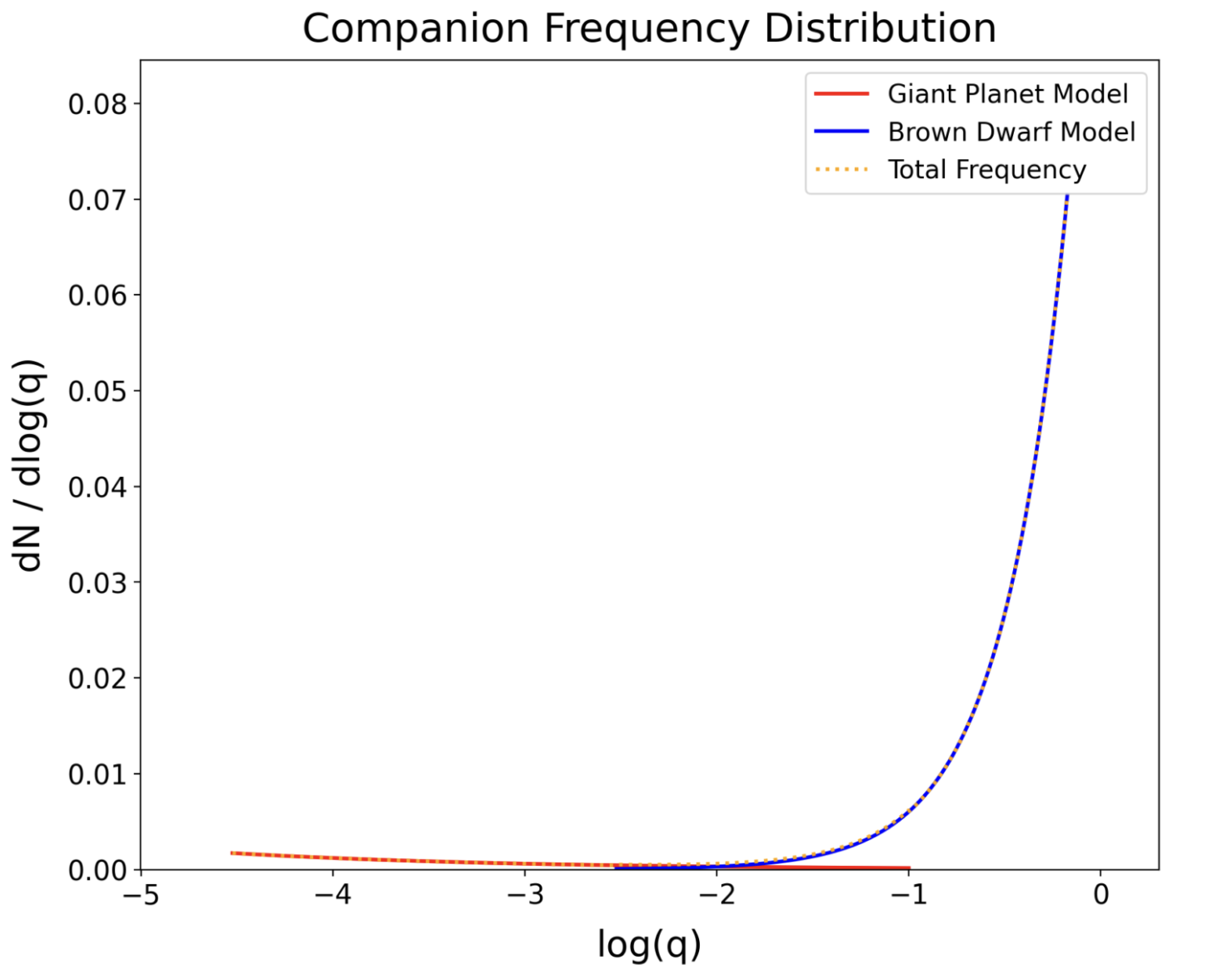} \label{fig:f11}}
      \hspace{0.5em}
      \subfloat[]{\includegraphics[width=0.3\textwidth]{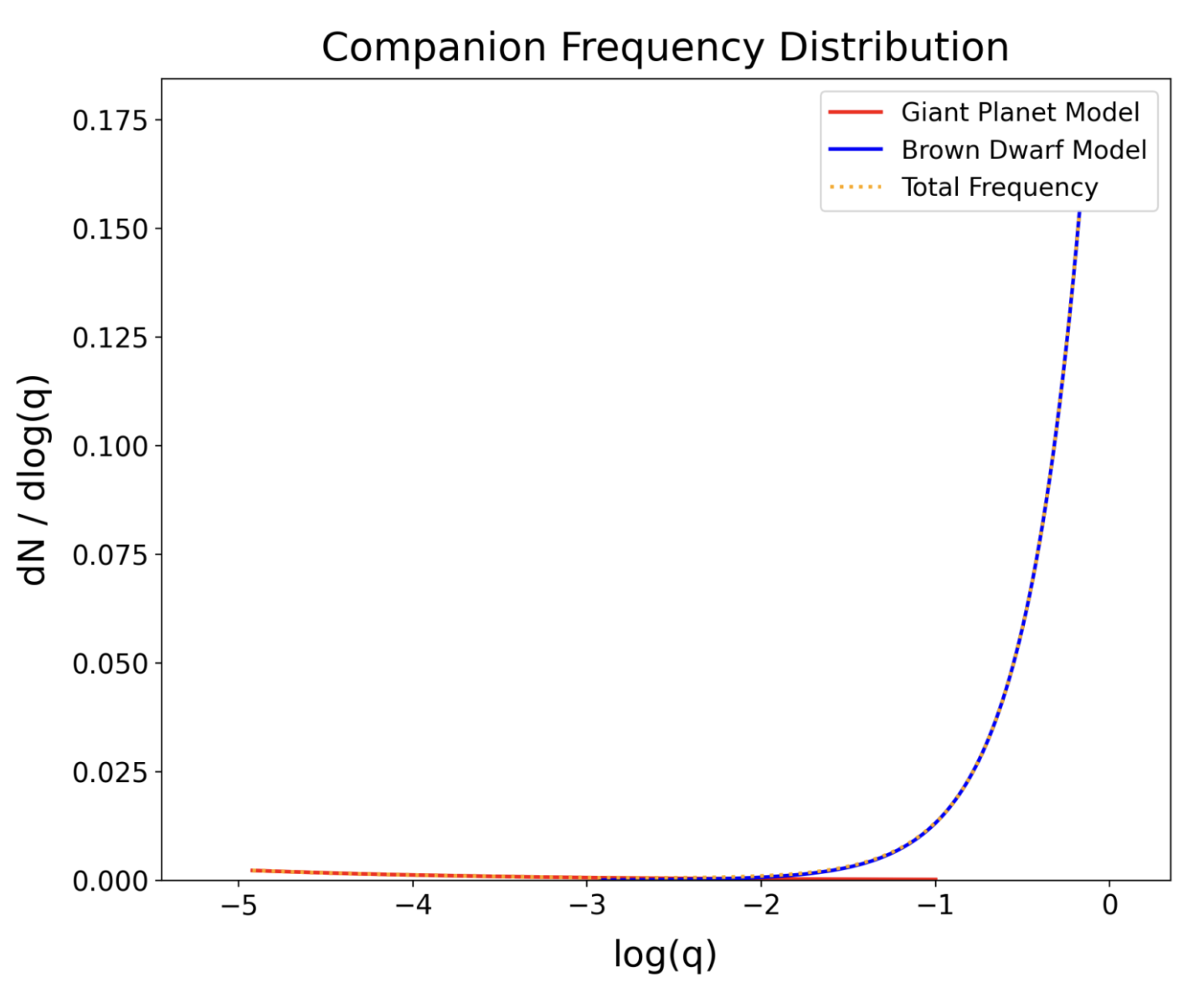} \label{fig:f12}}
      \caption{Companion mass ratio distribution from $-5 < log(q) < -0.2$ for M dwarf (left), FGK stars (middle), and A stars (right) for orbital separations from 100-1000 AU.  For all panels, brown dwarf binary distributions are in blue, the planet distributions are in red, and orange is the sum.}
      \label{Fig. 6}
    \end{figure*}
     
    Planet populations should dominate below the minimum mass considered for fragmentation, which we take here to be approximately 3 Jupiter masses, leading to an (in principle) observable discontinuity below this limit (cf. \citep{2018ApJ...853...37S}).  In these models the brown dwarf frequencies are generally so low, the discontinuity is difficult to see.  Similarly, where the model predicts a significant number of massive planets up to a planetary mass cut-off (some fraction of 10 \% the mass of the host star, the maximum for a stable circumstellar disk), there is another discontinuity that could in principle be observed.  For M dwarfs, there are significant numbers of brown dwarf companions predicted by the model below the 13 Jupiter mass deuterium burning limit, particularly beyond 10 AU where the binary companion distribution peaks.  There are also significant numbers of planetary mass companions above the 13 Jupiter mass deuterium burning limit around A stars, highlighting the problematic nature of this artificial boundary from the point of view of formation theory. \par

    \subsection{Mean Numbers of Companions per Star}
    
    Finally, we comment on the amplitudes derived in our model fitting which impact the number of companions expected over a given range of separation and mass ratio.  Our fits demonstrate that the brown dwarf contributions to very low 
    mass companions as for M, FGK, and A stars are consistent with simple extrapolations of the stellar companion distributions.  However, the two key fit parameters for the brown dwarf term in our model, $A_{bd}$ and $\beta$ are not well constrained.  It is oft stated that the binary frequency is much higher around stars of higher mass \citep[e.g.][]{2013ARA&A..51..269D}.  This is surely true if one only considers a stellar binary where both components must be above the hydrogen burning limit \citep{2006ApJ...640L..63L}.  However, if one takes the observed binary companion frequency over all separations as a function of host star mass, but only compare them over common ranges in q, the differences in binary frequency are not so stark (\cite{2022A&A...657A..48S}).   Of course, if the true limit to very low mass binary companions is the opacity limit for fragmentation at 3 Jupiter masses, there is naturally a greater fraction of multiplicity for higher mass stars which naturally span a larger range in q.  From this discussion, it is clear that companion frequency is only meaningful 
    over specified ranges of orbital separation and mass ratio.  \par

    What about the planet population?  From our best fit model, when we predict the integrated frequency of companions over all separations and over the same range of q, we expect comparable planet companion frequencies.  If we restrict the planet population to 1--10 Jupiter mass, this corresponds to different ranges of q for different host star mass such that, integrated over all separations, A/FGK/M stars are predicted to have frequencies of {\bf $ 0.151$, $0.115$, and $0.08$} respectively (mean numbers of planetary companions per star).  If predictions were extended to 0.3 Jupiter mass, these frequencies roughly double.  However, if we restrict planet populations to orbital separations beyond 30 AU, we are left with only 0.05 times these frequencies.  This corresponds to 0.006  for the mean number of planetary companion between 1--10 Jupiter mass beyond 30 AU for FGK stars, requiring a sample size of hundreds to insure a detection.   Our model is only intended to provide constraints on the frequency, orbital separation distribution, and planet mass function of gas giants.  If the results only obtain down to some fixed physical limit (e.g. $< 0.1$ Jupiter mass), then higher mass stars would naturally have more gas giant companions over a larger range of q implied by this lower limit.  
    
\section{Discussion}

Here we compare our results to others in the literature, as well as discuss important caveats of our work.  Finally, we discuss the implications of these results for star and planet formation theory. 

\subsection{Comparison to Other Results}

    How do predictions of our best fit model compare to independent analyses?  
      We compare our estimates to the frequency of brown dwarf companions from direct imaging surveys, over q ranges where they dominate.  \cite{2016PASP..128j2001B} provides point estimates of brown dwarf companions around stars as a function of host star mass and orbital separation based on a meta-analysis (“survey of surveys”) and our calculations agree with their results within the errors. 
    Local minima in the companion mass ratio distribution have been observed for FGK stars in the combined radial velocity and astrometric study of \cite{2011A&A...525A..95S} (cf. \citet{2019A&A...631A.125K}).   They find a local minimum in the companion mass function from 1-10 AU at 10s of Jupiter masses, consistent with the prediction of our model.    This local minimum is commonly referred to as the “brown dwarf desert” (most easily seen in Figures 5--6).  However, given the relatively flat q distribution for binaries, plotting the CMRD in log space simply means there is a very small log(q) range for brown dwarfs compared to stellar mass binaries.  A local minimum is also observed from microlensing observations (presumably dominated by M dwarf hosts) between -2.1 < log(q) < -1.4 (\cite{2016MNRAS.457.4089S}) probing the Einstein ring between 1-10 AU.  This range of q for 0.3 Msun primaries corresponds close to our estimate shown in Figure 5. \par

    As shown in Figure 2, this model explains the observed frequency of gas giants and brown dwarfs as a function of star mass reported in recent high contrast imaging surveys (\citep{2019AJ....158...13N}; \cite{Vigan_2021}, both used in the fitting).  For companions between 1-75 Jupiter mass observed from < 10 AU to > 100 AU, gas giants are more commonly detected around A stars than lower mass stars.  According to our model this is because this range corresponds to smaller q for intermediate mass stars compared to lower mass stars for a power-law planet mass function that rises to smaller q.  And for the opposite reason, brown dwarf companions are more common around M dwarfs over a fixed mass range (say 30-50 Jupiter masses) because they correspond to a higher q range for a flat mass function (rising in the log). \par

\subsection{Important Caveats}

 We reiterate that the brown dwarf part of our model is not well constrained with available data.  Future high contrast imaging surveys, as a function of host 
    star mass (e.g. the BEAST survey as well as the Y dwarf survey with JWST) will help test our model.  In particular, Gaia will provide comprehensive catalogues of sub--stellar companions around nearby stars 
    as a function of host star mass which will be crucial to better assessing whether is, or is not, a suppression of brown dwarf companions compared to extrapolations of stellar companion distributions. 
     
     Future work should calculate the likelihood of model parameters on an observation by observation basis rather than ``binning" inherent in using the point estimates from various surveys \citep[c.f.][]{2016ApJ...819..125C}.  There are other caveats to our approach that deserve attention.  Perhaps the planet mass function does indeed depend on orbital separation.  Recent results from the California Planet Survey suggests no dependence of the planet mass function above 30 Mearth masses on orbital separation out to a few AU \citep{2021ApJS..255...14F}.  Yet, there is evidence that the companion mass ratio distribution of multiple star systems depends on orbital radius for massive stars (\cite{2017ApJS..230...15M}).  More work is needed to address this issue.   
    
    Further, we have not fully treated the statistics of triple and higher order systems in our analysis.  Considering that primaries with a brown dwarf companion are binaries, we might wonder whether our fits apply to all primaries or only primaries without a {\it stellar} companion.  Perhaps systems with a wide orbit brown dwarf companion are less likely to have a gas giant.  We are only now beginning to find primaries with both a brown dwarf and a planet (\cite{2024A&A...689A.235R}). Perhaps systems without a stellar companion are more likely than random to have a brown dwarf companion.  It is interesting to note that many triple systems consist of a tight (a $<$ 100 days) stellar binary with a lower mass wide orbit tertiary.  There is no clear evidence that mass ratios in hierarchical systems are drawn from different distributions (\cite{2014MNRAS.437.1216D}; \cite{2010ApJS..190....1R}).   We do not have enough data here to address these questions.  While triple and higher order systems are rare (e.g. \cite{2019AJ....157..216W}; \cite{2010ApJS..190....1R}) they should be accounted for explicitly in future work.  We also ignore possible suppression formation in multiple star 
    systems (\cite{2016AJ....152....8K, 2021MNRAS.507.3593M}.  Finally, we also ignore possible correlations between planet properties in multi-planet systems (e.g. \cite{2024ApJ...968L..25B}; \cite{2019MNRAS.490.4575H}; \cite{2020ApJ...893..122D, 2016ApJ...819...83W, Bryan_2016} which should also be considered in the future. \par

\subsection{Implications for Theory}

    How do our results impact the development of theories of star formation?  Brown dwarf companions can be formed via turbulent fragmentation in molecular cloud cores \citep[e.g.][]{2010ApJ...725.1485O}, fragmentation in rotating flattened structures \citep[e.g.][]{2016ARA&A..54..271K}, and perhaps through classical gravitational instability in Keplerian disks (e.g. \cite{2013MNRAS.432.3168F}).  
    
    Different channels of binary formation may dominate as a function of separation and host star properties.  For example, companions within 1 AU may have mass ratios closer to unity (\cite{2020MNRAS.494.2289A} and references therein) as well as be preferentially formed in low metallicity systems \citep[cf][]{2018ApJ...854...44M}.  As more data are collected, hopefully we can discern details of how the brown dwarf companion mass ratio distribution correlates with orbital radius and host star mass.  This should enable us to differentiate the dominant modes of multiple formation.  What we can say for now is that a companion mass function that is universal in the mass ratio, q, independent of host star mass, is consistent with the data well into the brown dwarf regime, particularly for separations $1 AU < a < 100 AU$.  \par

    How do our results confront planet formation theory?  We find no evidence that the planet mass function depends on orbital separation which is somewhat surprising.  It is intriguing that the peak in the gas giant orbital separation distribution is close to location of the water-ice line, where we expect strong discontinuities in solid mass surface density as well as pressure, in the circumstellar disk.  Assuming $L \sim M^2$ in the pre-main sequence phase of evolution, and radiative equilibrium for blackbody grains ($T \sim L^{0.25}  \sim \sqrt{a}$), the ice-line should be proportional to the square--root of stellar mass.  We did find evidence for a trend in orbital separation peak with host star mass roughly consistent with this (1.7 AU at 0.3 M$_{\odot}$ versus 4.7 AU for 
    2.5 M$_{\odot}$).  More data are needed before the difference in orbital distributions versus host star mass can be considered robust.  Other models based on a planetesimal growth time as a function of disk properties (such as mass surface density and lifetimes) which in turn depend on stellar properties, predict orbital radii for core formation that does not depend strongly on stellar mass. For gravitational instability, given how mass surface density and the angular velocity depend on stellar mass, we expect the location where the Toomre Q instability parameter to drop below 1 to vary by only a factor of two in radius over a factor of six in stellar mass.  This discussion completely neglects the impact of orbital migration (Type I, II, and perhaps III) which is known to have an impact on the final orbital configurations of gas giant planets (\cite{2018haex.bookE.139N}).  Our model explains the frequency of gas giants observed as a function of stellar mass through a planet mass function that is a power-law in q, the ratio of planet to star mass.  Perhaps planet masses are dictated by circumstellar disk accretion rates, the lifetimes of disks, and additional random variables representing the microphysics of planet assembly.  If so, the form of the power-law, as well as upper and lower limits may be dictated by core formation times, as well as disk accretion rates and lifetimes as a function of stellar mass (\cite{2021ApJ...909....1A}).  We find no evidence that gas giant planet formation efficiency depends on stellar mass given our preferred model.   
    
    Our results only apply to gas giant planet formation.  At some point below a limit between 10--100 Earth masses, we anticipate that major planet populations form through another mechanism.  Assuming that disks around Sun--like stars begin forming their lasting generation of planets in disks approximately 10 \% the mass of the star, and assuming a gas to dust ratio of 100, there are initially about 300 Earth masses of solids from which to form planets.  In our own Solar System, we can account for about 100 Earth masses of solids in the known bodies suggesting planet formation is somewhat efficient.  Given that the typical Sun--like star does not form a gas giant (mean number of planetary companions per star above 0.3 Jupiter masses is about 20 \%) it is interesting to speculate on what sets rocky planet formation efficiency. Indeed low mass M dwarfs seem to have greater numbers of rocky planets compared to higher mass stars.  \cite{2015ApJ...814..130M} has shown that this is not merely a conservation of solids argument as low mass stars seem to turn solids into planets more efficiently than higher mass stars.  Yet gas giants might be an important part of what makes a system habitable.  It has been suggested that the presence of a gas giant could help separate the planet-forming reservoirs, making the formation of water worlds less likely (\cite{2021A&A...649L...5B}).  Preliminary work has even uncovered a tentative link between the presence of gas giants near the ice--line and temperate small planets (\cite{2024ApJ...968L..25B}).  If roughly 10 \% of Sun--like stars have gas giants near the ice--line, and 10 \% of these same stars have rocky planets in the habitable zone, and these properties are correlated, perhaps the frequency of habitable worlds around Sun--like stars is closer to 10 \% than the product of two independent factors (e.g. 1 \%). \par

    We have treated the multiple system brown dwarf companion distributions separately from a gas giant planetary population and found that we can fit all available data.  Models of gravitational instability indeed form companions with characteristic masses of 5--10 Jupiter masses beyond 30 AU (\cite{2013MNRAS.432.3168F}).  In our treatment, we would consider these objects part of the brown dwarf companion population.  Perhaps in fitting additional observations in the future, we will need to treat this population as distinct from an extension of the population of stellar mass binaries.  The planet population in this model has a power--law mass function rising toward masses as low as Saturn, and an orbital distribution peaked near 4 AU.  It is extremely tempting to associate this population with 
    the core accretion paradigm of gas giant planet formation. 
    
    \subsection{Future Work}
    
    Additional adaptive-optics imaging as a function of host star mass with existing 6-12 meter telescopes would be extremely valuable.  For example, the B star survey of \cite{2021arXiv210102043J} would probe predictions of the model at the highest stellar masses.  There already appears to be an upper limit to gas giant planet formation around 3 solar masses based on RV surveys of post--main sequence intermediate mass stars (\cite{2015A&A...574A.116R}).   This could be due to B stars losing the race to form the core of a gas giant before the gas disk dissipates (Adams et al. 2021).  Yet the BEAST survey has uncovered a curious 
    population of 5-20 Jupiter mass companions beyond 100 AU (Janson et al. 2021; other references).  Perhaps this population is due to a unique circumstance 
    capable of forming companions through GI around massive stars (Adams et al. 
    submitted). 
    
    Further, targeted imaging surveys of nearby faint host stars in the background limit with JWST will also yield crucial demographics for low mass ratios (e.g. \cite{2023ApJ...947L..30C}) for brown dwarf primaries.  Updated occurrence rates from radial velocity surveys of M dwarfs would also be useful.  Additional investigation of the mass distribution of microlensing host stars would also be welcome.  Gas giant occurrence rates for A stars from 1-10 AU are vital to better constrain the peak of the log-normal orbital distribution.  Astrometric results from Gaia will make significant contributions to demographics for all stars in the coming years (cf. \cite{2018A&A...614A..30R}).  Continuation of legacy duration radial velocity surveys (10+ years) will be crucial to bridge the gap with astrometry and direct imaging (cf. \citep{2016ApJ...817..104R, 2010Sci...330..653H}. The NASA/ESA/CSA James Webb Space Telescope has the sensitivity to detect ice giants beyond 10 AU around nearby young stars (Bogat et al. in preparation).  As ALMA observations of possibly planet-driven structure beyond 10 AU in several systems (\cite{2024A&A...685A..53G}), not to mention Neptune and Uranus in our own Solar System, suggest that the orbital distribution of ice giants may extend to larger separations than that observed for gas giants.  Surveys are underway with JWST to test this hypothesis. Finally,  
    future ELTs will have the spatial resolution to close the gap between direct imaging and radial velocity (\cite{2015IJAsB..14..279Q}). Indeed, combining all techniques on coherent star samples (RV, astrometry, direct imaging) yields a whole which is far greater than the sum of the parts  \citep[e.g.][]{2016ApJ...831..136C, 2019A&A...630A..50B, 2011A&A...525A..95S, 2018NatAs...2..883S}, \citep{2020arXiv201104703G}.  \par

\section{Summary}
    We have attempted to build a model of sub-stellar companions that considers both contributions from a gas giant planet population as well as very low mass brown dwarf multiple companions, as a function of orbital separation and host star mass.  This model is consistent with a wide variety of companion frequency estimates in the literature.  Our main conclusions are as follows: 
    
    \begin{enumerate}
        \item Our preferred model requires that both brown dwarf companions to stars, as well as the gas giant planet mass function are universal in the mass ratio of the companion to the host star, assuming that both the planet mass function and the multiple star companion mass ratio distribution do not depend on orbital separation (consistent with available data). 
        \item Our model predicts specific local minima in the companion mass ratio distribution as a function of orbital separation and host star mass between $0.001 < q < 0.1$, as well as possible discontinuities below the minimum mass for opacity-limited fragmentation for multiples and above the maximum planet mass predicted as a function of star mass.
        \item The brown dwarf desert could be a natural consequence of a flat companion mass ratio distribution for binary stars and the relative contribution of planets as a function of stellar mass. 
        \item Brown dwarf companions can be formed below the deuterium burning limit and gas giant planets can form above this limit.
        \item The peak in the orbital distribution of gas giants is consistent with 
        that expected from iceline considerations.  We find preliminary evidence that the peak is greater for higher mass (and luminosity) stars. 
        \item We see no strong trend in the efficiency of gas giant planet formation as a function of stellar mass, over a fixed range of q, integrated over all orbital separations.
        \item We provide estimates of gas giant planet frequencies as a function of survey sensitivity and orbital separation which depend on host star mass and an on--line calculator to facilitate making such predictions. 
    \end{enumerate}

\section*{Acknowledgements}
We are grateful to members of the FEPS Research Group over many years, in particular Abigail Guillait, Avery Peterson, Nicholas Susemiehl, Veenu Suri, Matthew De Furio, and Arthur Adams, as well as Arthur Vigan, Clemmence Fontanive, Fred Adams, Chris Miller, John Monnier, Tyler Gardner, and members of the NASA ExoPAG Science Interest Group 2 (Exoplanet Demographics) for helpful discussions. We are also grateful to Maddalena Reggiani, Sascha Quanz, Christoph Mordasini, and other members of the Swiss National Center of Competence in Research (PlanetS) whose past collaboration helped inspire this project. This work was partially supported by the University of Michigan and the NASA JWST NIRCam project (contract number NAS5-02105).


\bibliographystyle{aa}
\bibliography{aanda}

\begin{thebibliography}{87}
\expandafter\ifx\csname natexlab\endcsname\relax\def\natexlab#1{#1}\fi

\bibitem[{{Adams} {et~al.}(2023){Adams}, {Meyer}, {Howe}, {Burningham},
  {Daemgen}, {Fortney}, {Line}, {Marley}, {Quanz}, \&
  {Todorov}}]{2023AJ....166..192A}
{Adams}, A.~D., {Meyer}, M.~R., {Howe}, A.~R., {et~al.} 2023, \aj, 166, 192

\bibitem[{{Adams} {et~al.}(2020){Adams}, {Batygin}, \&
  {Bloch}}]{2020MNRAS.494.2289A}
{Adams}, F.~C., {Batygin}, K., \& {Bloch}, A.~M. 2020, \mnras, 494, 2289

\bibitem[{{Adams} {et~al.}(2021){Adams}, {Meyer}, \&
  {Adams}}]{2021ApJ...909....1A}
{Adams}, F.~C., {Meyer}, M.~R., \& {Adams}, A.~D. 2021, \apj, 909, 1

\bibitem[{{Adams} {et~al.}(1989){Adams}, {Ruden}, \&
  {Shu}}]{1989ApJ...347..959A}
{Adams}, F.~C., {Ruden}, S.~P., \& {Shu}, F.~H. 1989, \apj, 347, 959

\bibitem[{{Benz} {et~al.}(2014){Benz}, {Ida}, {Alibert}, {Lin}, \&
  {Mordasini}}]{2014prpl.conf..691B}
{Benz}, W., {Ida}, S., {Alibert}, Y., {Lin}, D., \& {Mordasini}, C. 2014, in
  Protostars and Planets VI, ed. H.~{Beuther}, R.~S. {Klessen}, C.~P.
  {Dullemond}, \& T.~{Henning}, 691

\bibitem[{{Bitsch} {et~al.}(2021){Bitsch}, {Raymond}, {Buchhave},
  {Bello-Arufe}, {Rathcke}, \& {Schneider}}]{2021A&A...649L...5B}
{Bitsch}, B., {Raymond}, S.~N., {Buchhave}, L.~A., {et~al.} 2021, \aap, 649, L5

\bibitem[{{Boehle} {et~al.}(2019){Boehle}, {Quanz}, {Lovis}, {S{\'e}gransan},
  {Udry}, \& {Apai}}]{2019A&A...630A..50B}
{Boehle}, A., {Quanz}, S.~P., {Lovis}, C., {et~al.} 2019, \aap, 630, A50

\bibitem[{{Bonfils} {et~al.}(2013){Bonfils}, {Delfosse}, {Udry}, {Forveille},
  {Mayor}, {Perrier}, {Bouchy}, {Gillon}, {Lovis}, {Pepe}, {Queloz}, {Santos},
  {S{\'e}gransan}, \& {Bertaux}}]{Bonfils_2013}
{Bonfils}, X., {Delfosse}, X., {Udry}, S., {et~al.} 2013, \aap, 549, A109

\bibitem[{{Borgniet} {et~al.}(2019){Borgniet}, {Lagrange}, {Meunier},
  {Galland}, {Arnold}, {Astudillo-Defru}, {Beuzit}, {Boisse}, {Bonfils},
  {Bouchy}, {Debondt}, {Deleuil}, {Delfosse}, {Desort}, {D{\'\i}az},
  {Eggenberger}, {Ehrenreich}, {Forveille}, {H{\'e}brard}, {Loeillet}, {Lovis},
  {Montagnier}, {Moutou}, {Pepe}, {Perrier}, {Pont}, {Queloz}, {Santerne},
  {Santos}, {S{\'e}gransan}, {da Silva}, {Sivan}, {Udry}, \&
  {Vidal-Madjar}}]{Borgniet_2019}
{Borgniet}, S., {Lagrange}, A.~M., {Meunier}, N., {et~al.} 2019, \aap, 621, A87

\bibitem[{{Bowler}(2016)}]{2016PASP..128j2001B}
{Bowler}, B.~P. 2016, \pasp, 128, 102001

\bibitem[{{Bowler} {et~al.}(2020){Bowler}, {Blunt}, \&
  {Nielsen}}]{2020AJ....159...63B}
{Bowler}, B.~P., {Blunt}, S.~C., \& {Nielsen}, E.~L. 2020, \aj, 159, 63

\bibitem[{{Bowler} {et~al.}(2015){Bowler}, {Liu}, {Shkolnik}, \&
  {Tamura}}]{Bowler_2015}
{Bowler}, B.~P., {Liu}, M.~C., {Shkolnik}, E.~L., \& {Tamura}, M. 2015, \apjs,
  216, 7

\bibitem[{{Bowler} {et~al.}(2023){Bowler}, {Tran}, {Zhang}, {Morgan}, {Ashok},
  {Blunt}, {Bryan}, {Evans}, {Franson}, {Huber}, {Nagpal}, {Wu}, \&
  {Zhou}}]{2023AJ....165..164B}
{Bowler}, B.~P., {Tran}, Q.~H., {Zhang}, Z., {et~al.} 2023, \aj, 165, 164

\bibitem[{{Bryan} {et~al.}(2016){Bryan}, {Bowler}, {Knutson}, {Kraus},
  {Hinkley}, {Mawet}, {Nielsen}, \& {Blunt}}]{Bryan_2016}
{Bryan}, M.~L., {Bowler}, B.~P., {Knutson}, H.~A., {et~al.} 2016, \apj, 827,
  100

\bibitem[{{Bryan} \& {Lee}(2024)}]{2024ApJ...968L..25B}
{Bryan}, M.~L. \& {Lee}, E.~J. 2024, \apjl, 968, L25

\bibitem[{{Buchner}(2016)}]{Buchner2016}
{Buchner}, J. 2016, {PyMultiNest: Python interface for MultiNest}, Astrophysics
  Source Code Library, record ascl:1606.005

\bibitem[{{Calissendorff} {et~al.}(2023){Calissendorff}, {De Furio}, {Meyer},
  {Albert}, {Aganze}, {Ali-Dib}, {Bardalez Gagliuffi}, {Baron}, {Beichman},
  {Burgasser}, {Cushing}, {Faherty}, {Fontanive}, {Gelino}, {Gizis},
  {Greenbaum}, {Kirkpatrick}, {Leggett}, {Martinache}, {Mary}, {N'Diaye},
  {Pope}, {Roellig}, {Sahlmann}, {Sivaramakrishnan}, {Thorngren}, {Ygouf}, \&
  {Vandal}}]{2023ApJ...947L..30C}
{Calissendorff}, P., {De Furio}, M., {Meyer}, M., {et~al.} 2023, \apjl, 947,
  L30

\bibitem[{{Clanton} \& {Gaudi}(2016)}]{2016ApJ...819..125C}
{Clanton}, C. \& {Gaudi}, B.~S. 2016, \apj, 819, 125

\bibitem[{{Crepp} {et~al.}(2016){Crepp}, {Gonzales}, {Bechter}, {Montet},
  {Johnson}, {Piskorz}, {Howard}, \& {Isaacson}}]{2016ApJ...831..136C}
{Crepp}, J.~R., {Gonzales}, E.~J., {Bechter}, E.~B., {et~al.} 2016, \apj, 831,
  136

\bibitem[{{Cumming} {et~al.}(2008){Cumming}, {Butler}, {Marcy}, {Vogt},
  {Wright}, \& {Fischer}}]{2008PASP..120..531C}
{Cumming}, A., {Butler}, R.~P., {Marcy}, G.~W., {et~al.} 2008, \pasp, 120, 531

\bibitem[{{De Furio} {et~al.}(2025){De Furio}, {Meyer}, {Greene}, {Hodapp},
  {Johnstone}, {Leisenring}, {Rieke}, {Robberto}, {Roellig}, {Cugno},
  {Fiorellino}, {Manara}, {Raileanu}, \& {van Terwisga}}]{2025ApJ...981L..34D}
{De Furio}, M., {Meyer}, M.~R., {Greene}, T., {et~al.} 2025, \apjl, 981, L34

\bibitem[{{De Rosa} {et~al.}(2014){De Rosa}, {Patience}, {Wilson}, {Schneider},
  {Wiktorowicz}, {Vigan}, {Marois}, {Song}, {Macintosh}, {Graham}, {Doyon},
  {Bessell}, {Thomas}, \& {Lai}}]{2014MNRAS.437.1216D}
{De Rosa}, R.~J., {Patience}, J., {Wilson}, P.~A., {et~al.} 2014, \mnras, 437,
  1216

\bibitem[{{Duch{\^e}ne} \& {Kraus}(2013)}]{2013ARA&A..51..269D}
{Duch{\^e}ne}, G. \& {Kraus}, A. 2013, \araa, 51, 269

\bibitem[{{Dulz} {et~al.}(2020){Dulz}, {Plavchan}, {Crepp}, {Stark}, {Morgan},
  {Kane}, {Newman}, {Matzko}, \& {Mulders}}]{2020ApJ...893..122D}
{Dulz}, S.~D., {Plavchan}, P., {Crepp}, J.~R., {et~al.} 2020, \apj, 893, 122

\bibitem[{{Duquennoy} \& {Mayor}(1991)}]{1991A&A...248..485D}
{Duquennoy}, A. \& {Mayor}, M. 1991, \aap, 500, 337

\bibitem[{{Fernandes} {et~al.}(2019){Fernandes}, {Mulders}, {Pascucci},
  {Mordasini}, \& {Emsenhuber}}]{2019ApJ...874...81F}
{Fernandes}, R.~B., {Mulders}, G.~D., {Pascucci}, I., {Mordasini}, C., \&
  {Emsenhuber}, A. 2019, \apj, 874, 81

\bibitem[{{Feroz} {et~al.}(2009){Feroz}, {Hobson}, \&
  {Bridges}}]{2009MNRAS.398.1601F}
{Feroz}, F., {Hobson}, M.~P., \& {Bridges}, M. 2009, \mnras, 398, 1601

\bibitem[{Feroz {et~al.}(2009)Feroz, Hobson, \& Bridges}]{Feroz_2009}
Feroz, F., Hobson, M.~P., \& Bridges, M. 2009, Monthly Notices of the Royal
  Astronomical Society, 398, 1601

\bibitem[{{Feroz} {et~al.}(2019){Feroz}, {Hobson}, {Cameron}, \&
  {Pettitt}}]{2019OJAp....2E..10F}
{Feroz}, F., {Hobson}, M.~P., {Cameron}, E., \& {Pettitt}, A.~N. 2019, The Open
  Journal of Astrophysics, 2, 10

\bibitem[{{Forgan} \& {Rice}(2013)}]{2013MNRAS.432.3168F}
{Forgan}, D. \& {Rice}, K. 2013, \mnras, 432, 3168

\bibitem[{{Fulton} {et~al.}(2021){Fulton}, {Rosenthal}, {Hirsch}, {Isaacson},
  {Howard}, {Dedrick}, {Sherstyuk}, {Blunt}, {Petigura}, {Knutson}, {Behmard},
  {Chontos}, {Crepp}, {Crossfield}, {Dalba}, {Fischer}, {Henry}, {Kane},
  {Kosiarek}, {Marcy}, {Rubenzahl}, {Weiss}, \& {Wright}}]{2021ApJS..255...14F}
{Fulton}, B.~J., {Rosenthal}, L.~J., {Hirsch}, L.~A., {et~al.} 2021, \apjs,
  255, 14

\bibitem[{{Garufi} {et~al.}(2024){Garufi}, {Ginski}, {van Holstein}, {Benisty},
  {Manara}, {P{\'e}rez}, {Pinilla}, {Ribas}, {Weber}, {Williams}, {Cieza},
  {Dominik}, {Facchini}, {Huang}, {Zurlo}, {Bae}, {Hagelberg}, {Henning},
  {Hogerheijde}, {Janson}, {M{\'e}nard}, {Messina}, {Meyer}, {Pinte}, {Quanz},
  {Rigliaco}, {Roccatagliata}, {Schmid}, {Szul{\'a}gyi}, {van Boekel},
  {Wahhaj}, {Antichi}, {Baruffolo}, \& {Moulin}}]{2024A&A...685A..53G}
{Garufi}, A., {Ginski}, C., {van Holstein}, R.~G., {et~al.} 2024, \aap, 685,
  A53

\bibitem[{{Gaudi} {et~al.}(2020){Gaudi}, {Christiansen}, \&
  {Meyer}}]{2020arXiv201104703G}
{Gaudi}, B.~S., {Christiansen}, J.~L., \& {Meyer}, M.~R. 2020, arXiv e-prints,
  arXiv:2011.04703

\bibitem[{{Grandjean} {et~al.}(2023){Grandjean}, {Lagrange}, {Meunier},
  {Chauvin}, {Borgniet}, {Desidera}, {Galland}, {Kiefer}, {Messina},
  {Iglesias}, {Nicholson}, {Pantoja}, {Rubini}, {Sedaghati}, {Sterzik}, \&
  {Zicher}}]{Grandjean_2023}
{Grandjean}, A., {Lagrange}, A.~M., {Meunier}, N., {et~al.} 2023, \aap, 669,
  A12

\bibitem[{{He} {et~al.}(2019){He}, {Ford}, \&
  {Ragozzine}}]{2019MNRAS.490.4575H}
{He}, M.~Y., {Ford}, E.~B., \& {Ragozzine}, D. 2019, \mnras, 490, 4575

\bibitem[{{Hirsch} {et~al.}(2021){Hirsch}, {Rosenthal}, {Fulton}, {Howard},
  {Ciardi}, {Marcy}, {Nielsen}, {Petigura}, {de Rosa}, {Isaacson}, {Weiss},
  {Sinukoff}, \& {Macintosh}}]{2021AJ....161..134H}
{Hirsch}, L.~A., {Rosenthal}, L., {Fulton}, B.~J., {et~al.} 2021, \aj, 161, 134

\bibitem[{{Howard} {et~al.}(2010){Howard}, {Marcy}, {Johnson}, {Fischer},
  {Wright}, {Isaacson}, {Valenti}, {Anderson}, {Lin}, \&
  {Ida}}]{2010Sci...330..653H}
{Howard}, A.~W., {Marcy}, G.~W., {Johnson}, J.~A., {et~al.} 2010, Science, 330,
  653

\bibitem[{{Janson} {et~al.}(2021){Janson}, {Squicciarini}, {Delorme},
  {Gratton}, {Bonnefoy}, {Reffert}, {Mamajek}, {Eriksson}, {Vigan}, {Langlois},
  {Engler}, {Chauvin}, {Desidera}, {Mayer}, {Marleau}, {Bohn}, {Samland},
  {Meyer}, {d'Orazi}, {Henning}, {Quanz}, {Kenworthy}, \&
  {Carson}}]{2021arXiv210102043J}
{Janson}, M., {Squicciarini}, V., {Delorme}, P., {et~al.} 2021, arXiv e-prints,
  arXiv:2101.02043

\bibitem[{{Kiefer} {et~al.}(2019){Kiefer}, {H{\'e}brard}, {Sahlmann}, {Sousa},
  {Forveille}, {Santos}, {Mayor}, {Deleuil}, {Wilson}, {Dalal}, {D{\'\i}az},
  {Henry}, {Hagelberg}, {Hobson}, {Demangeon}, {Bourrier}, {Delfosse},
  {Arnold}, {Astudillo-Defru}, {Beuzit}, {Boisse}, {Bonfils}, {Borgniet},
  {Bouchy}, {Courcol}, {Ehrenreich}, {Hara}, {Lagrange}, {Lovis}, {Montagnier},
  {Moutou}, {Pepe}, {Perrier}, {Rey}, {Santerne}, {S{\'e}gransan}, {Udry}, \&
  {Vidal-Madjar}}]{2019A&A...631A.125K}
{Kiefer}, F., {H{\'e}brard}, G., {Sahlmann}, J., {et~al.} 2019, \aap, 631, A125

\bibitem[{{Kratter} \& {Lodato}(2016)}]{2016ARA&A..54..271K}
{Kratter}, K. \& {Lodato}, G. 2016, \araa, 54, 271

\bibitem[{{Kraus} {et~al.}(2016){Kraus}, {Ireland}, {Huber}, {Mann}, \&
  {Dupuy}}]{2016AJ....152....8K}
{Kraus}, A.~L., {Ireland}, M.~J., {Huber}, D., {Mann}, A.~W., \& {Dupuy}, T.~J.
  2016, \aj, 152, 8

\bibitem[{{Lada}(2006)}]{2006ApJ...640L..63L}
{Lada}, C.~J. 2006, \apjl, 640, L63

\bibitem[{{Lannier} {et~al.}(2016){Lannier}, {Delorme}, {Lagrange}, {Borgniet},
  {Rameau}, {Schlieder}, {Gagn{\'e}}, {Bonavita}, {Malo}, {Chauvin},
  {Bonnefoy}, \& {Girard}}]{Lannier_2016}
{Lannier}, J., {Delorme}, P., {Lagrange}, A.~M., {et~al.} 2016, \aap, 596, A83

\bibitem[{{Madhusudhan}(2019)}]{2019ARA&A..57..617M}
{Madhusudhan}, N. 2019, \araa, 57, 617

\bibitem[{{McCarthy} \& {Zuckerman}(2004)}]{McCarthy_2004}
{McCarthy}, C. \& {Zuckerman}, B. 2004, \aj, 127, 2871

\bibitem[{{Metchev} \& {Hillenbrand}(2009)}]{Metchev_2009}
{Metchev}, S.~A. \& {Hillenbrand}, L.~A. 2009, \apjs, 181, 62

\bibitem[{{Meyer} {et~al.}(2018){Meyer}, {Amara}, {Reggiani}, \&
  {Quanz}}]{2018A&A...612L...3M}
{Meyer}, M.~R., {Amara}, A., {Reggiani}, M., \& {Quanz}, S.~P. 2018, \aap, 612,
  L3

\bibitem[{{Moe} \& {Di Stefano}(2017)}]{2017ApJS..230...15M}
{Moe}, M. \& {Di Stefano}, R. 2017, \apjs, 230, 15

\bibitem[{{Moe} \& {Kratter}(2018)}]{2018ApJ...854...44M}
{Moe}, M. \& {Kratter}, K.~M. 2018, \apj, 854, 44

\bibitem[{{Moe} \& {Kratter}(2021)}]{2021MNRAS.507.3593M}
{Moe}, M. \& {Kratter}, K.~M. 2021, \mnras, 507, 3593

\bibitem[{{Montet} {et~al.}(2014){Montet}, {Crepp}, {Johnson}, {Howard}, \&
  {Marcy}}]{Montet_2014}
{Montet}, B.~T., {Crepp}, J.~R., {Johnson}, J.~A., {Howard}, A.~W., \& {Marcy},
  G.~W. 2014, \apj, 781, 28

\bibitem[{{Mulders} {et~al.}(2015){Mulders}, {Pascucci}, \&
  {Apai}}]{2015ApJ...814..130M}
{Mulders}, G.~D., {Pascucci}, I., \& {Apai}, D. 2015, \apj, 814, 130

\bibitem[{{M{\"u}ller} {et~al.}(2018){M{\"u}ller}, {Helled}, \&
  {Mayer}}]{2018ApJ...854..112M}
{M{\"u}ller}, S., {Helled}, R., \& {Mayer}, L. 2018, \apj, 854, 112

\bibitem[{{Nelson}(2018)}]{2018haex.bookE.139N}
{Nelson}, R.~P. 2018, {Planetary Migration in Protoplanetary Disks}, ed. H.~J.
  {Deeg} \& J.~A. {Belmonte}, 139

\bibitem[{{Nielsen} {et~al.}(2019{\natexlab{a}}){Nielsen}, {De Rosa},
  {Macintosh}, {Wang}, {Ruffio}, {Chiang}, {Marley}, {Saumon}, {Savransky},
  {Ammons}, {Bailey}, {Barman}, {Blain}, {Bulger}, {Burrows}, {Chilcote},
  {Cotten}, {Czekala}, {Doyon}, {Duch{\^e}ne}, {Esposito}, {Fabrycky},
  {Fitzgerald}, {Follette}, {Fortney}, {Gerard}, {Goodsell}, {Graham},
  {Greenbaum}, {Hibon}, {Hinkley}, {Hirsch}, {Hom}, {Hung}, {Dawson},
  {Ingraham}, {Kalas}, {Konopacky}, {Larkin}, {Lee}, {Lin}, {Maire}, {Marchis},
  {Marois}, {Metchev}, {Millar-Blanchaer}, {Morzinski}, {Oppenheimer},
  {Palmer}, {Patience}, {Perrin}, {Poyneer}, {Pueyo}, {Rafikov}, {Rajan},
  {Rameau}, {Rantakyr{\"o}}, {Ren}, {Schneider}, {Sivaramakrishnan}, {Song},
  {Soummer}, {Tallis}, {Thomas}, {Ward-Duong}, \& {Wolff}}]{Nielsen_2019}
{Nielsen}, E.~L., {De Rosa}, R.~J., {Macintosh}, B., {et~al.}
  2019{\natexlab{a}}, \aj, 158, 13

\bibitem[{{Nielsen} {et~al.}(2019{\natexlab{b}}){Nielsen}, {De Rosa},
  {Macintosh}, {Wang}, {Ruffio}, {Chiang}, {Marley}, {Saumon}, {Savransky},
  {Ammons}, {Bailey}, {Barman}, {Blain}, {Bulger}, {Burrows}, {Chilcote},
  {Cotten}, {Czekala}, {Doyon}, {Duch{\^e}ne}, {Esposito}, {Fabrycky},
  {Fitzgerald}, {Follette}, {Fortney}, {Gerard}, {Goodsell}, {Graham},
  {Greenbaum}, {Hibon}, {Hinkley}, {Hirsch}, {Hom}, {Hung}, {Dawson},
  {Ingraham}, {Kalas}, {Konopacky}, {Larkin}, {Lee}, {Lin}, {Maire}, {Marchis},
  {Marois}, {Metchev}, {Millar-Blanchaer}, {Morzinski}, {Oppenheimer},
  {Palmer}, {Patience}, {Perrin}, {Poyneer}, {Pueyo}, {Rafikov}, {Rajan},
  {Rameau}, {Rantakyr{\"o}}, {Ren}, {Schneider}, {Sivaramakrishnan}, {Song},
  {Soummer}, {Tallis}, {Thomas}, {Ward-Duong}, \&
  {Wolff}}]{2019AJ....158...13N}
{Nielsen}, E.~L., {De Rosa}, R.~J., {Macintosh}, B., {et~al.}
  2019{\natexlab{b}}, \aj, 158, 13

\bibitem[{{Offner} {et~al.}(2010){Offner}, {Kratter}, {Matzner}, {Krumholz}, \&
  {Klein}}]{2010ApJ...725.1485O}
{Offner}, S. S.~R., {Kratter}, K.~M., {Matzner}, C.~D., {Krumholz}, M.~R., \&
  {Klein}, R.~I. 2010, \apj, 725, 1485

\bibitem[{{Petigura} {et~al.}(2013){Petigura}, {Marcy}, \&
  {Howard}}]{2013ApJ...770...69P}
{Petigura}, E.~A., {Marcy}, G.~W., \& {Howard}, A.~W. 2013, \apj, 770, 69

\bibitem[{{Quanz} {et~al.}(2015){Quanz}, {Crossfield}, {Meyer}, {Schmalzl}, \&
  {Held}}]{2015IJAsB..14..279Q}
{Quanz}, S.~P., {Crossfield}, I., {Meyer}, M.~R., {Schmalzl}, E., \& {Held}, J.
  2015, International Journal of Astrobiology, 14, 279

\bibitem[{{Raghavan} {et~al.}(2010){Raghavan}, {McAlister}, {Henry}, {Latham},
  {Marcy}, {Mason}, {Gies}, {White}, \& {ten Brummelaar}}]{2010ApJS..190....1R}
{Raghavan}, D., {McAlister}, H.~A., {Henry}, T.~J., {et~al.} 2010, \apjs, 190,
  1

\bibitem[{{Rameau} {et~al.}(2013){Rameau}, {Chauvin}, {Lagrange}, {Klahr},
  {Bonnefoy}, {Mordasini}, {Bonavita}, {Desidera}, {Dumas}, \&
  {Girard}}]{Rameau_2013}
{Rameau}, J., {Chauvin}, G., {Lagrange}, A.~M., {et~al.} 2013, \aap, 553, A60

\bibitem[{{Ranalli} {et~al.}(2018){Ranalli}, {Hobbs}, \&
  {Lindegren}}]{2018A&A...614A..30R}
{Ranalli}, P., {Hobbs}, D., \& {Lindegren}, L. 2018, \aap, 614, A30

\bibitem[{{Reffert} {et~al.}(2015){Reffert}, {Bergmann}, {Quirrenbach},
  {Trifonov}, \& {K{\"u}nstler}}]{2015A&A...574A.116R}
{Reffert}, S., {Bergmann}, C., {Quirrenbach}, A., {Trifonov}, T., \&
  {K{\"u}nstler}, A. 2015, \aap, 574, A116

\bibitem[{{Reggiani} \& {Meyer}(2013)}]{2013A&A...553A.124R}
{Reggiani}, M. \& {Meyer}, M.~R. 2013, \aap, 553, A124

\bibitem[{{Reggiani} {et~al.}(2016){Reggiani}, {Meyer}, {Chauvin}, {Vigan},
  {Quanz}, {Biller}, {Bonavita}, {Desidera}, {Delorme}, {Hagelberg}, {Maire},
  {Boccaletti}, {Beuzit}, {Buenzli}, {Carson}, {Covino}, {Feldt}, {Girard},
  {Gratton}, {Henning}, {Kasper}, {Lagrange}, {Mesa}, {Messina}, {Montagnier},
  {Mordasini}, {Mouillet}, {Schlieder}, {Segransan}, {Thalmann}, \&
  {Zurlo}}]{2016A&A...586A.147R}
{Reggiani}, M., {Meyer}, M.~R., {Chauvin}, G., {et~al.} 2016, \aap, 586, A147

\bibitem[{{Rosenthal} {et~al.}(2021){Rosenthal}, {Fulton}, {Hirsch},
  {Isaacson}, {Howard}, {Dedrick}, {Sherstyuk}, {Blunt}, {Petigura}, {Knutson},
  {Behmard}, {Chontos}, {Crepp}, {Crossfield}, {Dalba}, {Fischer}, {Henry},
  {Kane}, {Kosiarek}, {Marcy}, {Rubenzahl}, {Weiss}, \&
  {Wright}}]{2021ApJS..255....8R}
{Rosenthal}, L.~J., {Fulton}, B.~J., {Hirsch}, L.~A., {et~al.} 2021, \apjs,
  255, 8

\bibitem[{{Rowan} {et~al.}(2016){Rowan}, {Meschiari}, {Laughlin}, {Vogt},
  {Butler}, {Burt}, {Wang}, {Holden}, {Hanson}, {Arriagada}, {Keiser}, {Teske},
  \& {Diaz}}]{2016ApJ...817..104R}
{Rowan}, D., {Meschiari}, S., {Laughlin}, G., {et~al.} 2016, \apj, 817, 104

\bibitem[{{Ruggieri} {et~al.}(2024){Ruggieri}, {Desidera}, {Sozzetti},
  {Marzari}, {Pinamonti}, {Gratton}, {Biazzo}, {D'Orazi}, {Malavolta}, {Mesa},
  {Claudi}, {Benatti}, {Bignamini}, {Cabona}, {Chauvin}, {Hagelberg},
  {Mancini}, {Mantovan}, {Molinaro}, {Nardiello}, {Scandariato}, {Vigan}, \&
  {Zingales}}]{2024A&A...689A.235R}
{Ruggieri}, A., {Desidera}, S., {Sozzetti}, A., {et~al.} 2024, \aap, 689, A235

\bibitem[{{Sabotta} {et~al.}(2021){Sabotta}, {Schlecker}, {Chaturvedi},
  {Guenther}, {Mu{\~n}oz Rodr{\'\i}guez}, {Mu{\~n}oz S{\'a}nchez}, {Caballero},
  {Shan}, {Reffert}, {Ribas}, {Reiners}, {Hatzes}, {Amado}, {Klahr}, {Morales},
  {Quirrenbach}, {Henning}, {Dreizler}, {Pall{\'e}}, {Perger}, {Azzaro},
  {Jeffers}, {Kaminski}, {K{\"u}rster}, {Lafarga}, {Montes}, {Passegger}, \&
  {Zechmeister}}]{Sabotta_2021}
{Sabotta}, S., {Schlecker}, M., {Chaturvedi}, P., {et~al.} 2021, \aap, 653,
  A114

\bibitem[{{Sahlmann} {et~al.}(2011){Sahlmann}, {S{\'e}gransan}, {Queloz},
  {Udry}, {Santos}, {Marmier}, {Mayor}, {Naef}, {Pepe}, \&
  {Zucker}}]{2011A&A...525A..95S}
{Sahlmann}, J., {S{\'e}gransan}, D., {Queloz}, D., {et~al.} 2011, \aap, 525,
  A95

\bibitem[{{Schlaufman}(2018)}]{2018ApJ...853...37S}
{Schlaufman}, K.~C. 2018, \apj, 853, 37

\bibitem[{{Shvartzvald} {et~al.}(2016){Shvartzvald}, {Maoz}, {Udalski}, {Sumi},
  {Friedmann}, {Kaspi}, {Poleski}, {Szyma{\'n}ski}, {Skowron}, {Koz{\l}owski},
  {Wyrzykowski}, {Mr{\'o}z}, {Pietrukowicz}, {Pietrzy{\'n}ski},
  {Soszy{\'n}ski}, {Ulaczyk}, {Abe}, {Barry}, {Bennett}, {Bhattacharya},
  {Bond}, {Freeman}, {Inayama}, {Itow}, {Koshimoto}, {Ling}, {Masuda}, {Fukui},
  {Matsubara}, {Muraki}, {Ohnishi}, {Rattenbury}, {Saito}, {Sullivan},
  {Suzuki}, {Tristram}, {Wakiyama}, \& {Yonehara}}]{2016MNRAS.457.4089S}
{Shvartzvald}, Y., {Maoz}, D., {Udalski}, A., {et~al.} 2016, \mnras, 457, 4089

\bibitem[{{Snellen} \& {Brown}(2018)}]{2018NatAs...2..883S}
{Snellen}, I.~A.~G. \& {Brown}, A.~G.~A. 2018, Nature Astronomy, 2, 883

\bibitem[{{Sousa} {et~al.}(2015){Sousa}, {Santos}, {Mortier}, {Tsantaki},
  {Adibekyan}, {Delgado Mena}, {Israelian}, {Rojas-Ayala}, \&
  {Neves}}]{2015A&A...576A..94S}
{Sousa}, S.~G., {Santos}, N.~C., {Mortier}, A., {et~al.} 2015, \aap, 576, A94

\bibitem[{{Spitzer}(1978)}]{1978ppim.book.....S}
{Spitzer}, L. 1978, {Physical processes in the interstellar medium}

\bibitem[{{Susemiehl} \& {Meyer}(2022)}]{2022A&A...657A..48S}
{Susemiehl}, N. \& {Meyer}, M.~R. 2022, \aap, 657, A48

\bibitem[{{Suzuki} {et~al.}(2016){Suzuki}, {Bennett}, {Sumi}, {Bond}, {Rogers},
  {Abe}, {Asakura}, {Bhattacharya}, {Donachie}, {Freeman}, {Fukui}, {Hirao},
  {Itow}, {Koshimoto}, {Li}, {Ling}, {Masuda}, {Matsubara}, {Muraki},
  {Nagakane}, {Onishi}, {Oyokawa}, {Rattenbury}, {Saito}, {Sharan}, {Shibai},
  {Sullivan}, {Tristram}, {Yonehara}, \& {MOA
  Collaboration}}]{2016ApJ...833..145S}
{Suzuki}, D., {Bennett}, D.~P., {Sumi}, T., {et~al.} 2016, \apj, 833, 145

\bibitem[{{Vigan} {et~al.}(2017){Vigan}, {Bonavita}, {Biller}, {Forgan},
  {Rice}, {Chauvin}, {Desidera}, {Meunier}, {Delorme}, {Schlieder}, {Bonnefoy},
  {Carson}, {Covino}, {Hagelberg}, {Henning}, {Janson}, {Lagrange}, {Quanz},
  {Zurlo}, {Beuzit}, {Boccaletti}, {Buenzli}, {Feldt}, {Girard}, {Gratton},
  {Kasper}, {Le Coroller}, {Mesa}, {Messina}, {Meyer}, {Montagnier},
  {Mordasini}, {Mouillet}, {Moutou}, {Reggiani}, {Segransan}, \&
  {Thalmann}}]{Vigan_2017}
{Vigan}, A., {Bonavita}, M., {Biller}, B., {et~al.} 2017, \aap, 603, A3

\bibitem[{{Vigan} {et~al.}(2021){Vigan}, {Fontanive}, {Meyer}, {Biller},
  {Bonavita}, {Feldt}, {Desidera}, {Marleau}, {Emsenhuber}, {Galicher}, {Rice},
  {Forgan}, {Mordasini}, {Gratton}, {Le Coroller}, {Maire}, {Cantalloube},
  {Chauvin}, {Cheetham}, {Hagelberg}, {Lagrange}, {Langlois}, {Bonnefoy},
  {Beuzit}, {Boccaletti}, {D'Orazi}, {Delorme}, {Dominik}, {Henning}, {Janson},
  {Lagadec}, {Lazzoni}, {Ligi}, {Menard}, {Mesa}, {Messina}, {Moutou},
  {M{\"u}ller}, {Perrot}, {Samland}, {Schmid}, {Schmidt}, {Sissa}, {Turatto},
  {Udry}, {Zurlo}, {Abe}, {Antichi}, {Asensio-Torres}, {Baruffolo}, {Baudoz},
  {Baudrand}, {Bazzon}, {Blanchard}, {Bohn}, {Brown Sevilla}, {Carbillet},
  {Carle}, {Cascone}, {Charton}, {Claudi}, {Costille}, {De Caprio},
  {Delboulb{\'e}}, {Dohlen}, {Engler}, {Fantinel}, {Feautrier}, {Fusco},
  {Gigan}, {Girard}, {Giro}, {Gisler}, {Gluck}, {Gry}, {Hubin}, {Hugot},
  {Jaquet}, {Kasper}, {Le Mignant}, {Llored}, {Madec}, {Magnard}, {Martinez},
  {Maurel}, {M{\"o}ller-Nilsson}, {Mouillet}, {Moulin}, {Orign{\'e}}, {Pavlov},
  {Perret}, {Petit}, {Pragt}, {Puget}, {Rabou}, {Ramos}, {Rickman}, {Rigal},
  {Rochat}, {Roelfsema}, {Rousset}, {Roux}, {Salasnich}, {Sauvage}, {Sevin},
  {Soenke}, {Stadler}, {Suarez}, {Wahhaj}, {Weber}, \& {Wildi}}]{Vigan_2021}
{Vigan}, A., {Fontanive}, C., {Meyer}, M., {et~al.} 2021, \aap, 651, A72

\bibitem[{{Vigan} {et~al.}(2012){Vigan}, {Patience}, {Marois}, {Bonavita}, {De
  Rosa}, {Macintosh}, {Song}, {Doyon}, {Zuckerman}, {Lafreni{\`e}re}, \&
  {Barman}}]{Vigan_2012}
{Vigan}, A., {Patience}, J., {Marois}, C., {et~al.} 2012, \aap, 544, A9

\bibitem[{{Wagner} {et~al.}(2022){Wagner}, {Apai}, {Kasper}, {McClure}, \&
  {Robberto}}]{Wagner_2022}
{Wagner}, K., {Apai}, D., {Kasper}, M., {McClure}, M., \& {Robberto}, M. 2022,
  \aj, 163, 80

\bibitem[{{Weiss} {et~al.}(2016){Weiss}, {Rogers}, {Isaacson}, {Agol}, {Marcy},
  {Rowe}, {Kipping}, {Fulton}, {Lissauer}, {Howard}, \&
  {Fabrycky}}]{2016ApJ...819...83W}
{Weiss}, L.~M., {Rogers}, L.~A., {Isaacson}, H.~T., {et~al.} 2016, \apj, 819,
  83

\bibitem[{{Winters} {et~al.}(2019){Winters}, {Henry}, {Jao}, {Subasavage},
  {Chatelain}, {Slatten}, {Riedel}, {Silverstein}, \&
  {Payne}}]{2019AJ....157..216W}
{Winters}, J.~G., {Henry}, T.~J., {Jao}, W.-C., {et~al.} 2019, \aj, 157, 216

\bibitem[{Wittenmyer {et~al.}(2019)Wittenmyer, Butler, Horner, Clark, Tinney,
  Carter, Wang, Johnson, \& Collins}]{Wittenmyer_2019}
Wittenmyer, R.~A., Butler, R.~P., Horner, J., {et~al.} 2019, Monthly Notices of
  the Royal Astronomical Society, 491, 5248

\bibitem[{{Wittenmyer} {et~al.}(2016){Wittenmyer}, {Butler}, {Tinney},
  {Horner}, {Carter}, {Wright}, {Jones}, {Bailey}, \&
  {O'Toole}}]{2016ApJ...819...28W}
{Wittenmyer}, R.~A., {Butler}, R.~P., {Tinney}, C.~G., {et~al.} 2016, \apj,
  819, 28

\bibitem[{{Wolthoff} {et~al.}(2022){Wolthoff}, {Reffert}, {Quirrenbach},
  {Jones}, {Wittenmyer}, \& {Jenkins}}]{Wolthoff_2022}
{Wolthoff}, V., {Reffert}, S., {Quirrenbach}, A., {et~al.} 2022, \aap, 661, A63

\bibitem[{{Zhu}(2022)}]{Zhu_2022}
{Zhu}, W. 2022, \aj, 164, 5

\end{thebibliography}

\end{document}